\documentclass[11pt]{article}

\usepackage[margin=2.5cm]{geometry}
\usepackage{graphicx}
\usepackage{amsmath,amssymb}
\usepackage{bm}
\usepackage{booktabs}
\usepackage{longtable}
\usepackage{siunitx}
\DeclareSIUnit{\cycle}{cyc}
\usepackage[font=small,labelfont=bf]{caption}
\usepackage[numbers,sort&compress]{natbib}
\usepackage{placeins}
\usepackage[colorlinks=true,linkcolor=blue,citecolor=blue,urlcolor=blue]{hyperref}

\newcommand{\figorbox}[1]{%
  \IfFileExists{#1}%
    {\includegraphics[width=\linewidth]{#1}}%
    {\setlength{\fboxsep}{0pt}%
     \fbox{\begin{minipage}[c][0.34\linewidth][c]{\dimexpr\linewidth-2\fboxrule\relax}%
       \centering\ttfamily\small \detokenize{#1}\\[4pt]\normalfont\itshape (figure pending)%
     \end{minipage}}}%
}

\newcommand{\Itar}{\mathcal{I}_{\mathrm{tar}}}
\newcommand{\Jtar}{\mathcal{J}_{\mathrm{tar}}}

\newcommand{\um}{\ensuremath{\mu}m}
\newcommand{\eminus}{\ensuremath{e^-_{\mathrm{model}}}}
\newcommand{\snote}[1]{\section*{Supplementary Note #1}}

\newcommand{\snLibrary}{1}
\newcommand{\snOperator}{2}
\newcommand{\snPrior}{3}
\newcommand{\snInfo}{4}
\newcommand{\snModes}{5}
\newcommand{\snOpt}{6}
\newcommand{\snBaselines}{7}
\newcommand{\snImaging}{8}
\newcommand{\snExtended}{9}
\newcommand{\snLimits}{10}

\title{\bfseries Information-optimized color metalenses for camera imaging}

\author{Hyoseok Park, Yeonsang Park$^{*}$\\[2pt]
\small Department of Physics, Chungnam National University, Daejeon, Republic of Korea\\
\small $^{*}$Corresponding author: \texttt{yeonsang.park@cnu.ac.kr}}
\date{}

\begin{document}
\maketitle

% =====================================================================
\begin{abstract}
\noindent
A metalens is conventionally built by prescribing a target phase and matching
each meta-atom to it, and making that phase achromatic across the visible band is
the main difficulty in color imaging. We remove the prescribed phase entirely.
The meta-atom width map is optimized directly against the information the raw,
noisy, mosaic-sampled camera measurement carries about the target image,
differentiated through Maxwell propagation, the color-filter mosaic, pixel
integration and sensor noise, with material, thickness and library held fixed.
On a single-layer silicon-nitride platform, starting from a hyperbolic design,
this raises delivered target information by \num{15.8}\% despite lower collected
charge, produces the most balanced set of channel responses at the sensor plane
relative to their respective axial maxima, and reduces blue-channel blur. A high-signal-to-noise held-out
reconstruction improves in spatial fidelity and color accuracy
(PSNR $+1.28$~dB, $\Delta E_{00}$ $-18\%$). The same width-only
optimization improves color across silicon-nitride, silicon-dioxide and
titanium-dioxide libraries.
\end{abstract}

% =====================================================================
\section{Introduction}

Flat metasurface optics replace curved glass with a subwavelength-patterned
layer, and after a decade of development the meta-atom is a well-characterized
building block. Visible-band dielectric
libraries\citep{khorasaninejad2016,arbabi2015,devlin2016,zhan2016},
CMOS-compatible silicon-nitride processing\citep{colburn2018sin} and large-area
apertures\citep{park2024,wirthsingh2025,li2022inverse} are now routine, and the
complex transmission of each element is characterized across the visible band. A
color metalens is conventionally built from such a library by assigning a target
phase profile to its members: a hyperbolic profile for single-band focusing, or a
dispersion-engineered achromat that spends the available group delay on focal
coincidence across the band\citep{chen2018,wang2018,shrestha2018}. That step fixes
a local optical property of each element against a global phase target. It does
not ask whether the same library, assigned differently, would better serve the
camera built from it, a camera that receives broadband light, mixes wavelengths
through a color-filter mosaic, integrates over pixels, samples and aliases, and
reads out under shot and read noise before delivering a color image.

The question is sharpest for a single layer, because a patterned layer of fixed
thickness stores only a bounded group delay. Focusing a broad band onto one plane
requires an aperture-edge delay that grows with aperture and bandwidth, so the
achromatic aperture--bandwidth product is capped\citep{presutti2020,engelberg2021},
increasingly so with field of view\citep{shastri2022}. Design within that finite
budget is an allocation of transfer rather than a route to a shared focus.
Extended-depth-of-focus optics trade sharpness for a more invertible coded
blur\citep{dowski1995,colburn2018metasurface,colburn2020quartic,huang2020edof},
end-to-end methods co-optimize the optic with a reconstruction or inference
network\citep{sitzmann2018,chang2018,lin2021,tseng2021neural,tseng2021differentiable,chakravarthula2023,yang2024curriculum},
and broadband modulation-transfer volume supplies a further optical
criterion\citep{froch2025}. Each answers a specific optical or computational
question, yet none states how informative the raw camera measurement is about the
desired image, which is the quantity a computational color camera finally
delivers.

\begin{figure}[t]
\centering
\includegraphics[width=\linewidth]{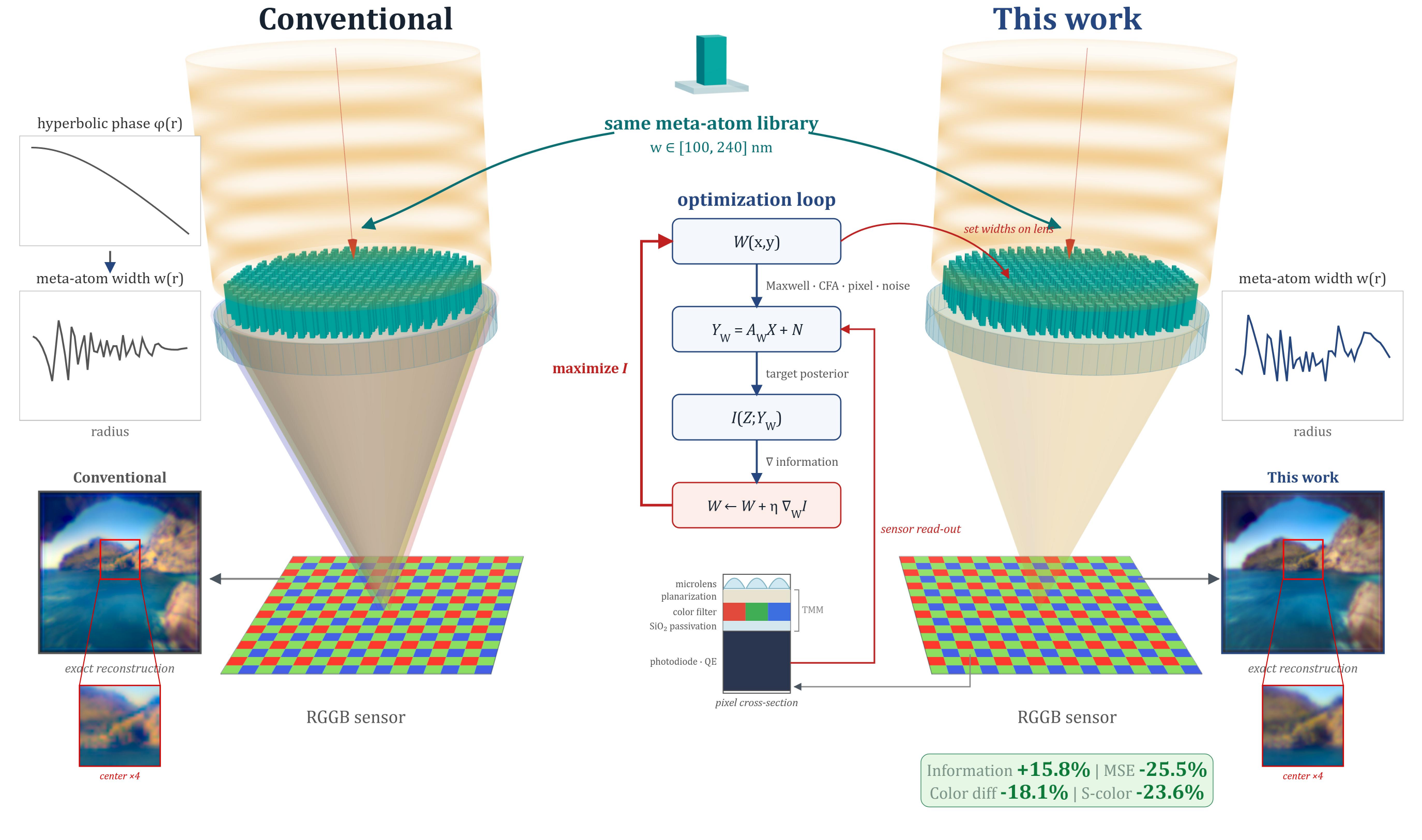}
\caption{\textbf{The missing step after meta-atom design.} Both routes draw
meta-atoms from an identical width library ($w\in[100,240]$~nm) and see the same
meta-atom library (local plane-wave response), focusing geometry and RGGB detector. \textbf{Left
(conventional):} a target hyperbolic phase $\varphi(r)$ is prescribed first and a
meta-atom width is matched to it at each radius, so the phase is the object of the
design and the width map $w(r)$ merely implements it, and the camera never enters the
design. \textbf{Right (this work):} no phase is prescribed. The width map
$W(x,y)$ is optimized directly against the information $I(Z;Y_W)$ that the
complete camera measurement $Y_W=A_W X+N$ carries about the desired image $Z=TX$,
differentiating through Maxwell propagation, the color-filter mosaic, pixel
integration, sampling and aliasing, and shot and read noise (central loop). A
conventional hyperbolic map only initializes the search. Because the material,
layer count, thickness, pitch and meta-atom library are identical in both routes,
only the objective the widths are optimized for changes. The direct-simulation
reconstructions (bottom, with central crops) improve by
$+15.8\%$ in information and $+1.28$~dB in PSNR (a $25.5\%$ reduction in
reconstruction mean-squared error) while the color error drops by
$18.1\%$ ($\Delta E_{00}$) and $23.6\%$ (S-CIELAB).}
\label{fig:workflow}
\end{figure}

Our contribution is to change what the width map is optimized for
(Fig.~\ref{fig:workflow}). Conventional color-metalens design prescribes a phase
profile and then matches a meta-atom to it at every site, so the phase is the
object of the design and the widths only implement it. We remove that
intermediate target. No desired phase enters the loop. No decoder enters it
either, so the optic is not co-optimized with, or tuned to, a particular
reconstruction network. The information the
complete camera measurement carries about the wanted image is differentiated
through the full physical chain and back onto the meta-atom width at each lattice
site, with a conventional hyperbolic map used only to initialize the search.
Everything else about the optic is held fixed. To our knowledge this combination
has not been reported for a color metalens. On the fixed
silicon-nitride platform studied here it raises the delivered target information
by \num{15.8}\% over the projected hyperbolic reference while the camera
collects less charge. The same width-only optimization improves delivered
information and reconstructed color on silicon-dioxide and titanium-dioxide
meta-atom libraries as well (Figs.~\ref{fig:optical} and~\ref{fig:imaging}), so the result is not
specific to one material.

The criterion by which the widths are reallocated is the information the camera
measurement carries about the desired image. Information-theoretic assessment of
optical systems has a long history\citep{difrancia1955,fellgett1955,cox1986} and
now spans task-specific imaging\citep{neifeld2003,ashok2008}, prefilters for
undersampled detection\citep{greivenkamp1990,euliss2001}, point-spread-function
design for parameter estimation\citep{shechtman2014}, information-driven
imaging-system design\citep{pinkard2024}, lensless encoders\citep{kabuli2026} and
phase-mask front ends for classification\citep{wang2026classification}. These
methods primarily maximize the mutual information about the object or the
noiseless sensor image, $I(X;Y)$\citep{pinkard2024,kabuli2026}. In a companion
study we optimize a nanophotonic color router, a color-filter-array replacement,
for the field-dependent camera pupil rather than for plane-wave
routing\citep{park2026colorrouter}, scoring it by the information the sensor
reading carries about the desired image. Here the free element is a focusing
metalens read out through an unchanged RGGB mosaic, and we set its width map by
the target information the raw, aliased frame carries about that image. For a color
camera the latent spectral scene $X$ is not the delivered product. The desired
image is a tristimulus map $Z=TX$. Because $Z$ is a deterministic function of
$X$, the information chain rule separates the captured information,
\begin{equation}
I(X;Y)=I(Z;Y)+I(X;Y\mid Z)\equiv I_{\mathrm{tar}}+I_{\mathrm{null}},
\label{eq:decomposition}
\end{equation}
into a delivered target term $I_{\mathrm{tar}}=I(Z;Y)$ and a target-null
remainder $I_{\mathrm{null}}=I(X;Y\mid Z)$ that records the metameric spectral
variation the color target discards. The design objective used here is the
delivered target term $I(Z;Y_W)$, a posterior quantity independent of any trained
decoder. Posterior information about a downstream
quantity of interest is a standard device of goal-oriented Bayesian experimental
design\citep{attia2018goed}, and task-specific information for detection and
classification targets is
established\citep{neifeld2003,neifeld2007tsi,ashok2008,wang2026classification}.
Covariance-aware and capacity objectives have recently been explored
for photonic and diffractive
systems\citep{chen2026incoherentmi,kienesberger2026metaimager,markley2025idealio}.
In those the latent source is itself the signal, whereas the target here is a
high-dimensional continuous full-color image delivered through spectral mixing,
pixel integration, sampling and aliasing. For a scene $X_\theta$, a
width-dependent camera measurement $Y_{W,\theta}$ and a desired image
$Z_\theta=TX_\theta$ at field point $\theta$, the design question is not which
wavelength focuses most sharply, but how much observing $Y_{W,\theta}$ reduces
uncertainty about $Z_\theta$.

We formalize that question as
\begin{equation}
X_\theta\longrightarrow
Y_{W,\theta}=A_{W,\theta}X_\theta+N_\theta,\qquad
Z_\theta=TX_\theta,
\label{eq:intro-channel}
\end{equation}
where $A_{W,\theta}$ contains wavelength-dependent propagation, calibrated
sensor throughput, pixel integration, color-filter response, sampling, aliasing
and field dependence, and $N_\theta$ contains shot and read noise. The target
$T$ declares the image the camera is intended to deliver. We then choose
\begin{equation}
W^\star=\arg\max_{W\in\mathcal F}\Itar(W),\qquad
\Itar(W)\equiv
\sum_{\theta\in\Theta}w_\theta
I\!\left(Y_{W,\theta};Z_\theta\right),
\qquad \sum_{\theta\in\Theta}w_\theta=1.
\label{eq:design-principle}
\end{equation}
This definition neither chooses a trained decoder nor substitutes a
reconstruction score for the channel. It measures camera-delivered target
information under an explicit scene prior, noise model and target.

We demonstrate the principle primarily on a fixed single-layer silicon-nitride
platform. Material, thickness, lattice, aperture, focal length and meta-atom
library are
held fixed, and only the spatial assignment of pillar widths is optimized from
the feasible-set projection of a hyperbolic map. Large-scale meta-optic inverse
design is established, and already optimizes individual meta-atom parameters
directly over large apertures\citep{pestourie2018,li2022inverse}. Those methods
target an optical figure such as focusing efficiency or a prescribed phase. The
contribution here is the objective and the way it reaches the widths: the
decoder-independent target information of a complete sampled, mosaicked and noisy
color camera is differentiated directly onto the meta-atom width at each site,
with no target phase imposed. A conventional RGGB detector is one concrete
realization of $A_W$, and the same posterior formulation applies whenever the
physical measurement and the target can be stated.

% =====================================================================
\section{Camera-level information design}

\subsection*{Camera model}

\begin{figure}[t]
\centering
\includegraphics[width=\linewidth]{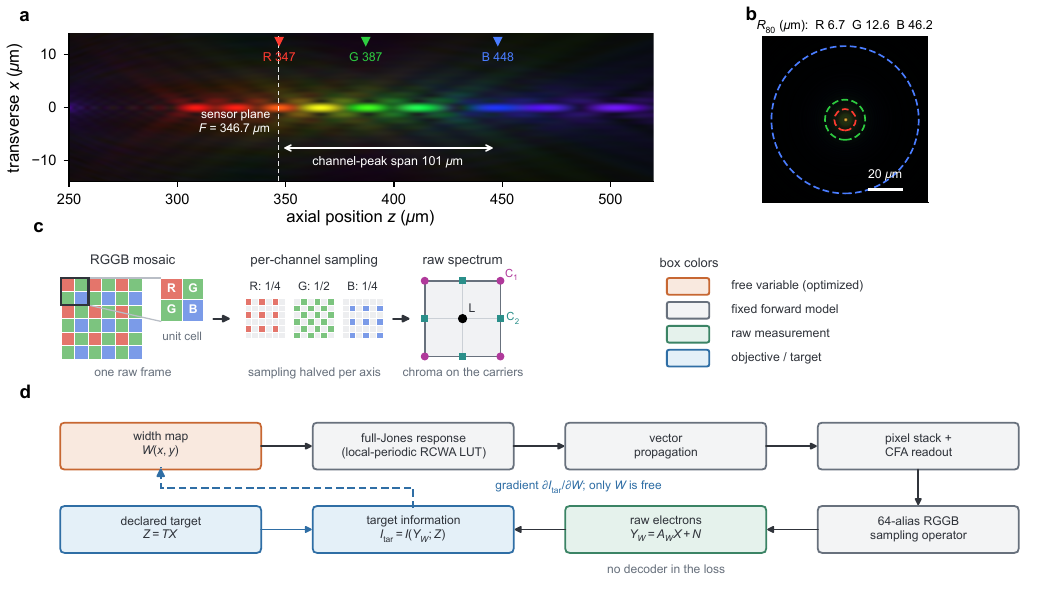}
\caption{\textbf{Chromatic focal structure, the mosaic measurement and the
single objective.}
\textbf{a} True-color longitudinal composite of the hyperbolic reference from
the saved $x$--$z$ diagnostic, per-wavelength peak-normalized for hue
visibility and not radiometric: the R, G and B color-filter channel peaks sit
at \num{347}, \num{387} and \SI{448}{\micro\meter}, a span of
\SI{101}{\micro\meter} (every peak interior to the saved
\SIrange{250}{520}{\micro\meter} window), while the sensor samples one plane,
$F=\SI{346.7}{\micro\meter}$.
\textbf{b} On-axis detector-plane PSF as a CFA-channel composite, with the dashed
circles marking the channel EE80 radii, \num{6.73}, \num{12.61} and
\SI{46.20}{\micro\meter} in R, G and B. \textbf{c} The RGGB mosaic subsamples
the color channels, R and B at one quarter of the sensor sites and G at one half,
and places chroma on spatial carriers (schematic).
\textbf{d} The differentiable chain from the width map to the raw measurement
$Y_W=A_WX+N$, scored against the target $Z=TX$ by the single objective
$\Itar=I(Y_W;Z)$. No decoder appears in the loss and the gradient returns only
to $W$ (schematic). Panels a and b are saved-data renders, and panels c and d
carry no numerical results.}
\label{fig:concept}
\end{figure}

The differentiable camera chain begins with the full-Jones RCWA forward
model and maps a scene to raw model-electron coordinates
(Fig.~\ref{fig:concept}).
An object point launches a spherical wave onto the pupil, each pillar applies a
complex transmission from the meta-atom response, the field propagates
to the sensor, and the pixel stack and readout convert irradiance to an
exposure-calibrated model-electron signal (Methods). After linearization about
the mean scene, the camera model is written at each field and spatial
frequency as
\begin{equation}
Y_{W,\theta}(\nu)=
A_{W,\theta}(\nu)X_\theta(\nu)+N_\theta(\nu).
\label{eq:measurement}
\end{equation}
The operator $A_{W,\theta}$ includes optical propagation, model-electron-domain
throughput under the exposure calibration, CFA response, pixel
integration, sampling and aliasing: the \SI{0.25}{\micro\meter}
scene-to-\SI{2.0}{\micro\meter} RGGB-cell chain folds \num{64} scene aliases
into each coarse frequency, and the exact multirate polyphase construction of
$A_{W,\theta}$ and of the \num{12}-row target operator is given in Supplementary
Note~\snOperator. Because each RGGB cell provides only four raw measurements,
the optic cannot independently preserve every color--spatial target direction,
and the design freedom determines which combinations receive the available
measurement signal-to-noise.

\subsection*{The target-information objective}

The primary desired image is full RGB on the \SI{1.0}{\micro\meter} target
lattice, defined by the target operator
$T_a(\nu)\in\mathbb C^{12\times64K}$ built from the linear-sRGB projection $T$
(Supplementary Note~\snOperator). For a Gaussian scene prior with covariance
$\mathbf{C}_{\mathrm{al}}$ and Gaussian-equivalent electron noise, the posterior
covariance of the latent scene is $P_{W,\theta}$, and the only design objective
is
\begin{equation}
\boxed{\displaystyle
\Itar(W)=
\sum_{\theta\in\Theta}w_\theta
I\!\left(Y_{W,\theta};Z_\theta\right)
=\sum_{\theta\in\Theta}w_\theta
\left\langle\tfrac14\cdot\tfrac12\log_2
\frac{\det\!\left(T_a\mathbf{C}_{\mathrm{al}}T_a^{\dagger}\right)}
     {\det\!\left(T_a P_{W,\theta}T_a^{\dagger}\right)}
\right\rangle_{\nu}}
\label{eq:targetinfo}
\end{equation}
in bit per raw pixel, where $\sum_\theta w_\theta=1$ and the factor $1/4$
converts one four-sample RGGB cell to a raw-pixel normalization. Thus
\begin{equation}
\Itar=
\sum_{\theta\in\Theta}w_\theta
\left[H(Z_\theta)-H(Z_\theta\mid Y_{W,\theta})\right],
\label{eq:entropy-reduction}
\end{equation}
and maximizing $\Itar$ minimizes the posterior volume of the target,
not a focal or reconstruction proxy. Throughput, spectral mixing, sampling,
aliasing and shot/read noise enter through the same model-electron-domain
operator. Under this stated linear--Gaussian model the conditional mean is a
sufficient statistic for $Z_\theta$, so Eq.~\eqref{eq:targetinfo} is a property
of the camera channel, prior and target, independent of any particular
demosaicker (Supplementary Note~\snInfo). Its scope is the linear--Gaussian image model.
Reconstruction is used only for post-optimization imaging checks.

\subsection*{Modal interpretation}

If $\gamma_i(\nu,\theta)$ denotes a target-relevant modal signal-to-noise
ratio, the local information per RGGB cell is
\begin{equation}
\Jtar(\nu,\theta)=
\tfrac12\sum_i\log_2[1+\gamma_i(\nu,\theta)],\qquad
\frac{\partial\Jtar}{\partial\gamma_i}
=\frac{1}{2\ln 2}\frac{1}{1+\gamma_i},
\label{eq:modal-information}
\end{equation}
and the reported objective is
$\Itar=\tfrac14\sum_\theta w_\theta
\langle\Jtar(\nu,\theta)\rangle_\nu$.
The marginal return is smaller for an already strong mode and larger for a weak
one. This concavity produces a tendency to reallocate finite optical transfer
toward weak target-relevant modes. The related I--MMSE
identity\citep{guo2005mmse} connects information to reconstruction utility over
signal-to-noise ratio. One objective therefore weighs throughput, spectral
mixing, the color-filter mosaic, pixel integration, aliasing, noise and the
scene prior together.

\subsection*{Connection to optical transfer proxies}

The relation to transfer proxies is explicit in a restricted scalar channel. If
$Y(\nu)=H_W(\nu)X(\nu)+N(\nu)$, then
\begin{equation}
I(X;Y)=\tfrac12\sum_\nu\log_2\!\left[
1+\frac{S_X(\nu)}{S_N(\nu)}|H_W(\nu)|^2\right].
\label{eq:scalar-information}
\end{equation}
With a white scene prior, white noise, one channel, no aliasing, no target
projection and the low-SNR approximation $\log(1+x)\simeq x$, this reduces to
$I(X;Y)\propto\sum_\nu|H_W(\nu)|^2$: squared transfer energy is recovered only
after the spectral coupling, noise, scene statistics, sampling and target
structure of the full camera have been removed. Broadband volume under the MTF
has been used as a published meta-optic design objective~\citep{froch2025}.
MTF measures transferred contrast, whereas $\Itar$ measures which measurement modes are
distinguishable and useful for estimating the color image. In particular,
scalar channelwise transfer retains only diagonal magnitudes: two color
operators can have the same diagonal transfer energy while their
target-conditioned Gram matrices are rank one (collinear responses, no color
discrimination) or full rank (independent color responses). The full derivation
of the restricted limit is given in Supplementary Note~\snModes.

% =====================================================================
\section{Fixed-platform width-only optimization}

Everything about the optic except the widths is fixed to the
single-layer silicon-nitride platform: SiN pillars of height \SI{1000}{nm} on a
\SI{290}{nm} square lattice, circular aperture $D=\SI{208}{\micro\meter}$,
focal distance $F=\SI{346.7}{\micro\meter}$, one meta-atom response library,
one sensor stack and one exposure calibration. The design variable is only the
spatial map $W(x,y)$. The exact corrected \SI{600}{nm} hyperbolic map, projected
onto the feasible set, is the optimizer's step-0 map. The implemented feasible
set $\mathcal F$ imposes the usable width interval by exact projection and the
mirror-quadrant symmetry, and nothing else: no spatial filter and no
smoothness term. The hyperbolic reference is evaluated from its stored map under
the same projection, so the design space and the reference are scored on the
same footing. For the information design, the optimizer, checkpoint selection and
final design selection use $-\Itar$ and no image-quality metric. The MTF-volume
control is selected by its own native MTF-volume objective on the same schedule.

The comparison set is three arms on identical physical degrees of freedom,
width library, spectral and field sampling, optimization budget and learning
rate: the hyperbolic reference, the information design and a matched-budget
broadband MTF-volume control motivated by published MTF-volume
maximization\citep{froch2025}. Every arm is seeded from the same projected
hyperbolic reference, so the native objective is the only intentional
difference between the arms. Each design is scored from its stored full width
map by one common evaluator at the fixed common exposure. A design with a lower
model-electron signal is penalized through the shot-noise covariance, so
brightness differences enter the comparison through the noise model itself.

% =====================================================================
\section{Results}

\subsection*{Information gain under an identical physical platform}

The saved 25-field evaluator orders the designs as follows
(Table~\ref{tab:main}). The information design delivers the highest weighted
$\Itar$, \num{0.5860} bit per raw pixel, the hyperbolic reference delivers
\num{0.5059} and the matched-budget MTF-volume control \num{0.4180}. Every arm
is seeded from the reference, so these values are also each design's change in
delivered target information from its own initialization: the information
design gains \num{15.8}\% and the MTF-volume control loses \num{17.4}\%. The
information design collects a weighted mean raw-pixel charge of \num{624.4}
model electrons, against \num{724.0} for the reference and \num{274.8} for the
MTF-volume control at the same fixed exposure, so it delivers more target
information than the reference from a smaller collected charge. The full
target/total/conditional decomposition is reported in Supplementary
Note~\snExtended.

\begin{table}[t]
\centering
\caption{\textbf{Delivered target information and collected charge at the fixed
common exposure.} 25-field quadrature-weighted values. The mean raw-pixel charge
is $(R+2G+B)/4$ on the same quadrature. Every design is scored by one evaluator
from its stored width map.}
\label{tab:main}
\begin{tabular}{lcc}
\hline
Design & $\Itar$ (bit/raw px) & charge (model-e\textsuperscript{-}/raw px)\\
\hline
Hyperbolic reference & 0.5059 & 724.0\\
MTF-volume control & 0.4180 & 274.8\\
Information design & 0.5860 & 624.4\\
\hline
\end{tabular}
\end{table}

\subsection*{Chromatic reallocation of optical transfer}

The matched-budget MTF and information maps each redraw the width assignment
across the aperture relative to the hyperbolic reference (Supplementary
Note~\snExtended), reshaping the per-channel spot sizes. The reference
is a red-focused hyperbolic, sharpest in the 600-nm band but blurring the blue
channel, and the information design gives up that 600-nm concentration for a
flatter chromatic profile across the band (Fig.~
\ref{fig:imaging}b).

Equation~\eqref{eq:targetinfo} names no wavelength, focus or spot size. Its
modal form instead assigns optical transfer according to marginal
target-information return (Eq.~\ref{eq:modal-information}): a strong mode can
contribute substantial absolute information while offering little return for
another increment of transfer, and the same increment can be more valuable in a
weak color-separation mode. In the optimized designs the reallocation
brings the per-channel spot sizes toward a common value. Across the visible
band the reference concentrates its focus in the 600-nm band and blurs the blue
channel, while the information design holds a flatter per-wavelength spot size
(Fig.~\ref{fig:imaging}b). The imbalance is intrinsic to the hyperbolic
prescription rather than to one focus choice.
The information design broadens its sharpest channel and tightens its broadest,
so the balancing concedes the reference's red-band concentration rather than
transporting it to the other channels. In the transfer domain the reference's
\SI{600}{nm} ridge at the \SI{0.21}{\cycle\per\micro\meter} cut falls from
\num{0.704} to \num{0.342} for the information design while the \num{540} and
\SI{570}{nm} values rise. Aggregated into color channels, the same reallocation
appears in the on-axis channel transfer (Fig.~\ref{fig:channelmtf}). The
reference concentrates transfer in the red channel and leaves the blue near zero,
the information design raises the blue and green channels and lowers the red so
the three channel transfers become comparable, and the MTF-volume control raises
the red without equalizing the channels. At the \SI{0.2}{\cycle\per\micro\meter}
cut the on-axis channel MTF moves from $(0.32,0.06,0.01)$ for R, G and B in the
reference to $(0.15,0.15,0.04)$ in the information design, matching the red and
green channels and lifting the blue, whereas the MTF-volume control reaches
$(0.23,0.04,0.05)$ and holds the green channel below the reference. The full wavelength-resolved transfer
maps are given in Supplementary Note~\snExtended. The three designs keep their per-channel
global intensity maxima spread over about \SI{100}{\micro\meter} axially, so no
single depth holds every channel's absolute peak. At the sensor plane itself
each channel nonetheless retains a local intensity value at or below that peak,
and the information design balances these sensor-plane values across channels
(\num{0.85}, \num{0.87} and \num{0.40} of each channel's own maximum for R, G
and B) more evenly than the reference (\num{1.00}, \num{0.52}, \num{0.17}) or
the MTF-volume control (\num{0.98}, \num{0.55}, \num{0.26}) (Supplementary
Note~\snExtended). The operative equalization therefore reaches the sensor plane,
where the channel responses relative to their respective axial maxima become the
most balanced set, as well as the channel spot sizes, even though the dominant
focal depth stays spread.

Because the objective is evaluated over the full field quadrature, this
reallocation is not restricted to the on-axis chromatic response. Field- and
wavelength-dependent aberrations enter through the same camera operator and are
penalized only insofar as they reduce target information. The optimization
therefore does not explicitly minimize chromatic or geometric aberration. It
redistributes optical transfer across wavelength and field according to their
effect on the sampled camera measurement.

The stored physical diagnostics and charge are recorded but not differentiated,
not used for checkpoint selection and not combined with the loss. The same common exposure applies to the direct signed-field evaluation: on
the $8\times8$ uniform-field-mean convention the per-channel $(R,G,B)$ electron
charges are $(546.3,\,771.0,\,270.0)$ for the reference,
$(457.9,\,699.9,\,266.3)$ for the information design and
$(284.9,\,413.4,\,159.0)$ for the MTF-volume control. On this convention the
information design collects \num{9.93}\% less charge than the reference, and
\num{13.75}\% less on the fixed 25-field quadrature weighting
(Table~\ref{tab:main}).

\begin{figure}[tbp]
\centering
\includegraphics[width=\linewidth]{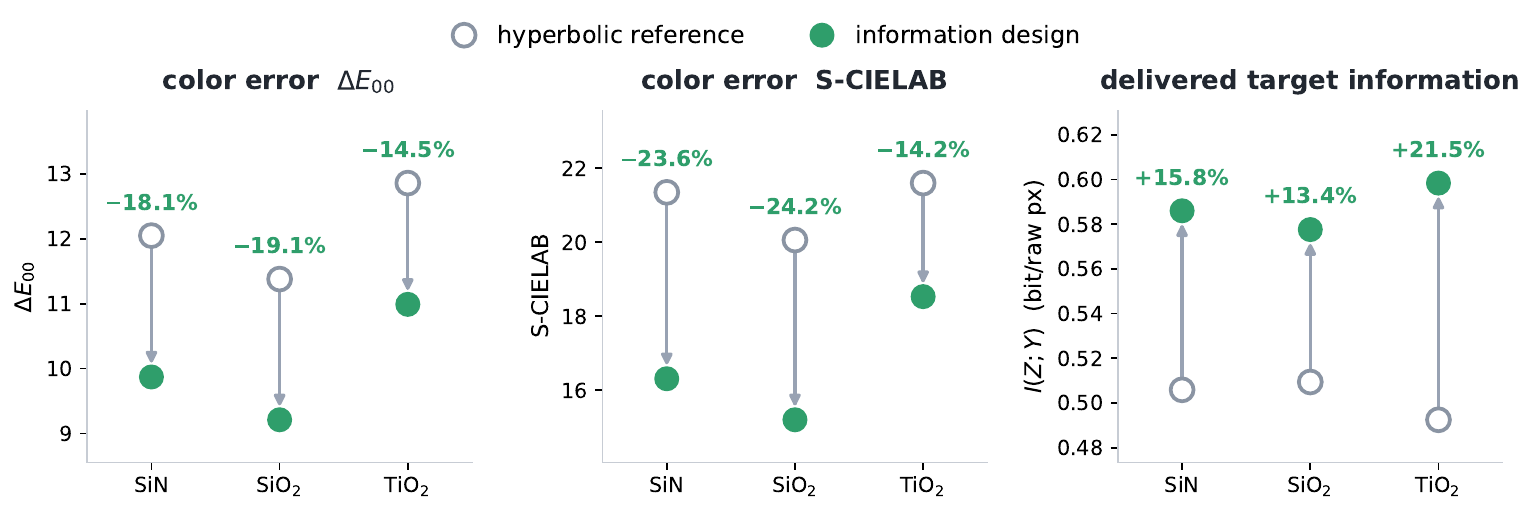}
\caption{\textbf{Color error and delivered information across meta-atom
materials.} In every panel the hyperbolic reference (open circle) and the
information design (filled circle) are joined for each material, with the
information design's relative change annotated. \textbf{a} Saved-image color
error $\Delta E_{00}$ and \textbf{b} S-CIELAB, each design reconstructed under
its own independent analytic color calibration, lower is better.
\textbf{c} Delivered target information $I(Z;Y)$,
higher is better (SiN values in Table~\ref{tab:main}). The information design lowers both
color-error measures and raises delivered information in SiN, SiO$_2$ and
TiO$_2$ alike.}
\label{fig:optical}
\end{figure}

\begin{figure}[tbp]
\centering
\includegraphics[width=\linewidth]{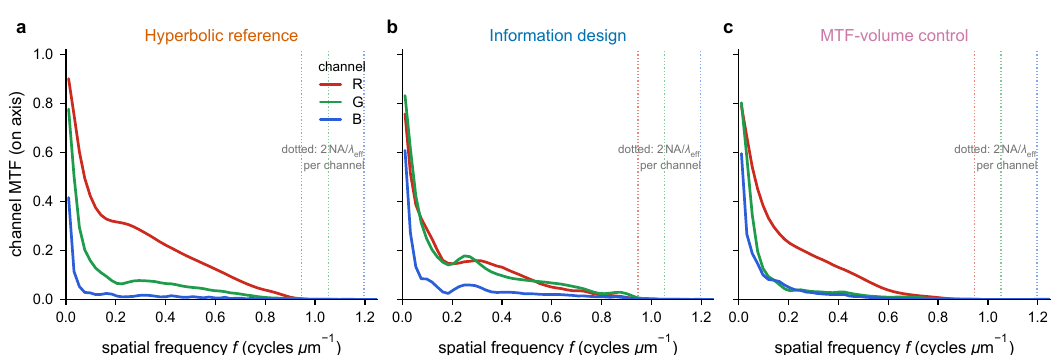}
\caption{\textbf{On-axis channel modulation transfer.} Polychromatic R, G and B
channel MTF at the sensor plane for the hyperbolic reference (\textbf{a}), the
information design (\textbf{b}) and the matched MTF-volume control (\textbf{c}).
Each channel is the radial average of the on-axis complex optical transfer
function at the nine design wavelengths weighted by the Samsung Galaxy S20
color-filter response. The hyperbolic reference concentrates transfer in the red
channel and leaves the blue channel near zero. The information design raises the
blue and green channels and lowers the red, so the three channel responses become
the most balanced set. The MTF-volume control raises the red channel without
equalizing the three. Dotted lines mark the per-channel diffraction cutoff
$2\,\mathrm{NA}/\lambda_\mathrm{eff}$ at the exact geometric air NA \num{0.287}.}
\label{fig:channelmtf}
\end{figure}

\subsection*{The matched MTF-volume control}

Broadband MTF-volume maximization has been proposed as a design criterion for
full-color meta-optic imaging\citep{froch2025}, and this control tests that
criterion directly under the present sampled color camera. The control is a
strong design on its own terms. Because every arm is seeded
from the hyperbolic reference, the seed value of the MTF-volume objective is the
reference's own MTF volume, so the recorded optimization stage is a direct
reference-versus-control comparison on the proxy: the DC-normalized MTF volume
rises from \num{0.0491} to \num{0.0711} in the frequency-area units, an
improvement of \num{44.6}\%. That control delivers \num{17.4}\% less target
information than the reference on the field-weighted aggregate, while the
information design sits \num{15.8}\% above it. Resolved by field, the control
falls below the reference in all \num{25} fields (Supplementary
Note~\snExtended). The MTF-volume objective is optimized successfully, but the
DC-normalized MTF-volume proxy is insufficient for this sampled color camera
under common exposure.

The reversal has several contributing causes, among them absolute throughput,
spectral channel collinearity, alias folding and target projection. The most
direct is that broadband MTF volume rewards the magnitude of
transferred contrast, taken one wavelength and one channel at a time, while the
target information rewards measurement modes that are distinguishable and useful
for estimating the color image. A DC-normalized, per-channel transfer magnitude
misses what the camera actually depends on. Two color responses can each be
large yet point in the same spectral direction, so the second adds contrast but
little new color information. The full target-conditioned matrix model
(Eq.~\ref{eq:targetinfo}, Supplementary Note~\snModes) separates a set of
collinear, rank-deficient responses from an independent one, whereas scalar
transfer magnitude does not.
The information is concave in modal signal-to-noise
(Eq.~\ref{eq:modal-information}), so strengthening an already strong mode
returns little while reviving a weak target-relevant mode returns much, and the
useful move is to raise the least distinguishable target modes rather than to
equalize channel spot sizes. Consistent with this distinction, the control reaches
the most uniform per-channel spot sizes of the three designs, yet delivers the
least target information (Table~\ref{tab:main}). Only the transfer that survives sampling carries
information, and the \SI{0.25}{\micro\meter} scene folds \num{64} aliases into
each coarse RGGB frequency, so contrast that aliases across channels lifts the
MTF volume without improving distinguishability, while a DC-normalized proxy
stays blind to the absolute charge and hence to shot noise. Computed in the
model-electron domain, the target information accounts for all of these jointly,
which is how the control improves the MTF volume by \num{44.6}\% while losing
\num{17.4}\% of the target information.

\subsection*{Image reconstruction and color metrics}

Image formation and metric evaluation are separated. The forward simulation
first stores raw model-electron and reconstruction arrays for the hyperbolic,
MTF-volume and information designs. Every image is formed by direct full-scene
propagation. No field is approximated by a mirrored or rotated field, and
geometric symmetries are used only for computational acceleration. A second,
forward-free step evaluates PSNR, CIEDE2000 color error and S-CIELAB from the
stored arrays. No imaging metric, scene or reconstruction result is available to
the optimizer.

The comparison uses one exposure calibrated on the reference and then fixed
across designs, because Eq.~\eqref{eq:targetinfo} already captures throughput and
shot noise under common exposure. White balance and color correction are fixed
by a per-design analytic calibration artifact before test scenes are rendered,
and each design uses its own fixed matrix. A same-scene, ground-truth-fitted
gain is an oracle and is not used.

\begin{figure}[tbp]
\centering
\includegraphics[width=0.80\linewidth]{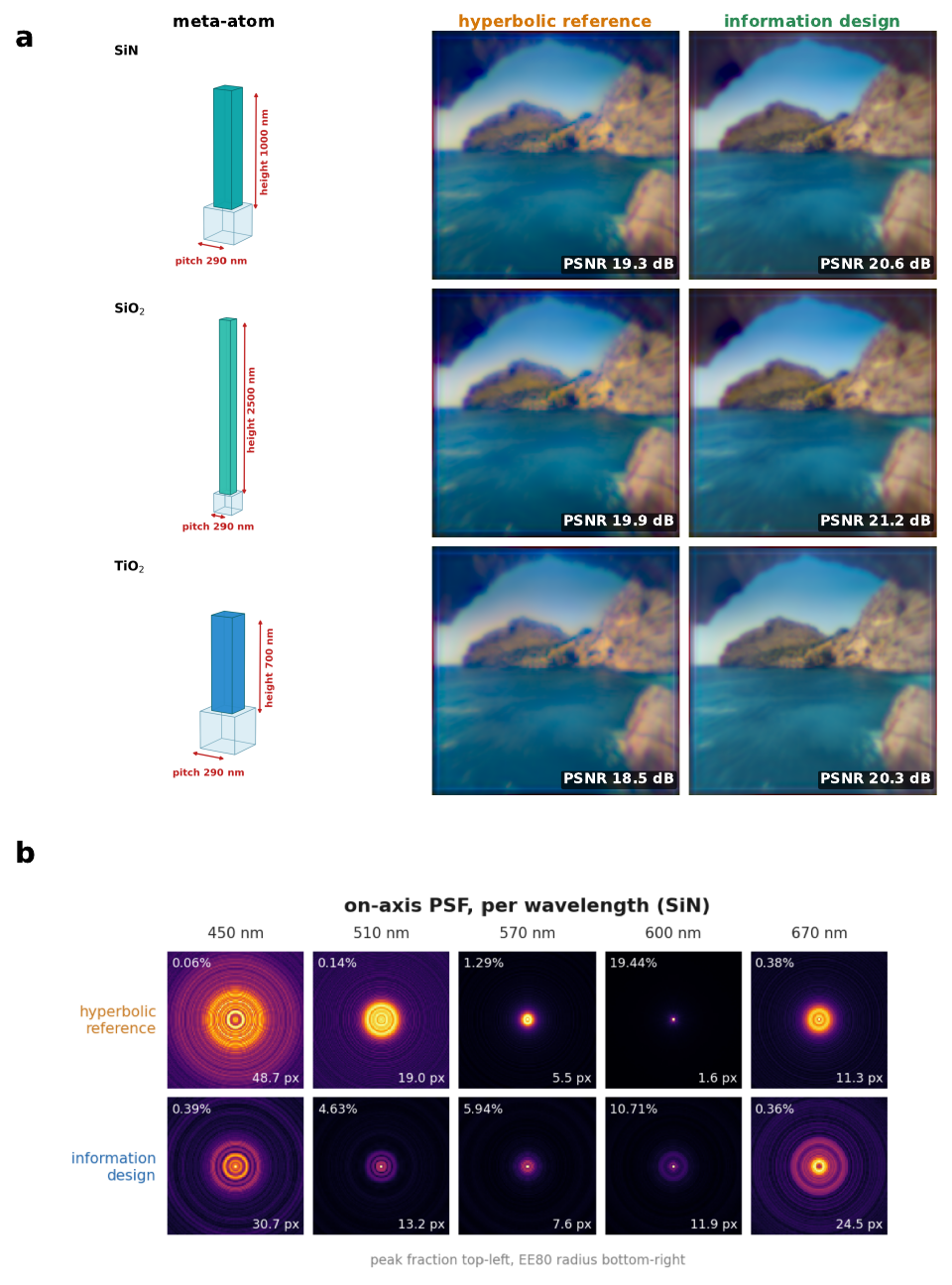}
\caption{\textbf{Material generality: width-only information optimization beats the
hyperbolic reference across SiN, SiO$_2$ and TiO$_2$.}
\textbf{a} Reconstructed landscape for each material (rows) under the
hyperbolic reference and the information design (columns), beside the
corresponding meta-atom geometry. Every image is formed by direct full-scene raw
propagation with bilinear demosaic and a fixed field-dependent Wiener step. Each
design is reconstructed under its own independent analytic color calibration,
and the field-of-view PSF bank enters only the reconstruction kernels and never
forms the image.
Each reconstruction carries its PSNR against the ground truth, and in every material
the information design raises PSNR and lowers color error (color error and
delivered information in Fig.~\ref{fig:optical}).
\textbf{b} Per-wavelength on-axis PSF for SiN, reference above and information
design below, each tile self-peak-normalized with the peak-pixel energy fraction
top left and the EE80 radius bottom right. The reference concentrates focus in
the 600-nm band (tightest at \SI{600}{nm}) while blurring the blue channel, and the
information design flattens the focus across the band.}
\label{fig:imaging}
\end{figure}

One held-out scene provides a post-optimization imaging check consistent with
the information ranking. In the high-SNR limit the information design
reconstructs this scene best: its PSNR, $\Delta E_{00}$ and S-CIELAB are
\SI{20.58}{\decibel}, \num{9.87} and \num{16.31}, against \SI{19.30}{\decibel},
\num{12.05} and \num{21.35} for the hyperbolic reference, each design under its
own analytic color calibration (Fig.~\ref{fig:imaging}, Fig.~\ref{fig:optical}a,b).
The \SI{1.28}{\decibel} PSNR gain is a \num{25.5}\% reduction in reconstruction
mean-squared error. That ranking holds across the exposures of ordinary
and bright photography and reverses only near \SI{15}{\decibel}
reference-equivalent signal-to-noise ratio, at a few tens of model-electrons per
pixel in the lower-exposure regime of the calibrated model, where the concavity
that rewards reallocation gives way to collected charge (Supplementary
Note~\snExtended).
The MTF-volume control is weakest on every axis.
Across seven further held-out natural scenes rendered through the identical
pipeline, the information design lowers both color-error measures on every scene
(mean $\Delta E_{00}$ \num{11.52}$\,\rightarrow\,$\num{9.53}, $-17\%$, and
S-CIELAB $-13\%$) and improves PSNR on five of the seven (mean $+0.30$~dB),
reported per scene in Supplementary Note~\snExtended and shown in
Supplementary Figs.~6 and~7.
Every image is formed by direct full-scene propagation: the
reconstruction takes the saved raw bank through demosaicing and a
field-dependent Wiener step, and the point-spread-function bank enters only as
the Wiener decoder.

The same phase-free width-only optimization applies to other meta-atom
materials. Repeated on silicon-dioxide (SiO$_2$) and titanium-dioxide (TiO$_2$)
libraries with the aperture, focal length, lattice, sensor and exposure fixed as
for silicon nitride (Supplementary Table~2), it reproduces the same ordering. On the held-out scene the
information design again raises reconstruction PSNR over the projected hyperbolic
reference, from \num{19.94} to \SI{21.24}{\decibel} for SiO$_2$ and \num{18.53}
to \SI{20.26}{\decibel} for TiO$_2$, matching the
\num{19.30}$\,\rightarrow\,$\SI{20.58}{\decibel} silicon-nitride gain, and in
each material it lowers both color-error measures and delivers more target
information (Fig.~\ref{fig:optical}, Fig.~\ref{fig:imaging}).

\FloatBarrier

\FloatBarrier
\subsection*{Independent full-wave validation of the spatial intensity structure}

The local-periodic forward model is verified against an independent full-wave
solver. For the exact $D=208$\,\si{\micro\meter} target-information width map,
we compared the forward model to Tidy3D finite-difference time-domain
simulation of the complete aperture, using the same centered
$D=208$\,\si{\micro\meter} circular pupil for both propagation paths. The two
agree closely across the nine design wavelengths, with a mean full-$x$--$z$
intensity cosine similarity of \num{0.9863} and an rms focal-position
difference of \SI{1.29}{\um} (Supplementary Note~\snLimits,
Supplementary Figs.~9 and~10). The comparison is on axis, after per-wavelength
normalization, and covers the longitudinal and transverse intensity, not the
full complex field or the off-axis camera operator.

% =====================================================================
\FloatBarrier
\section{Discussion}

Once a manufacturable library exists, the
conventional hyperbolic phase assignment is not in general the optimum of the
camera built from it, and the freedom that remains, which member of the same
library sits at each lattice site, can be set by the information the physical
measurement provides about the desired image. Once the scene ensemble, detector,
noise, sampling and target are specified, they induce the objective instead of
adding a term to a lens-only score.
Equation~\eqref{eq:targetinfo} implements the statement directly, so the width
map changes the camera channel and the objective measures the resulting reduction
in target uncertainty. PSF, MTF, charge, focal position and image metrics keep
their diagnostic roles, but none is an additional design objective.

The same distinction applies to optical aberrations. Because $\Itar$ is
evaluated over the field of view, wavelength- and field-dependent aberrations
enter the optimization through their effect on the raw camera measurement rather
than through separate aberration penalties. The optimizer is therefore free to
suppress aberrations that strongly reduce target information while retaining
others whose removal would provide little information gain. In the present
design, for example, the channel spot sizes become more balanced and the channel
responses at the sensor plane become the most balanced of the three designs
relative to their respective axial maxima, with blue-channel blur reduced, even though the chromatic focal
spread remains large, showing that information optimization need not coincide with
conventional achromatization. Rather than prescribing which
aberrations to minimize, the information objective weights their consequences
through the complete camera measurement.

The physical interpretation is allocation under a finite dispersion and transfer
budget. The marginal information return falls as
a mode's signal-to-noise rises, so the optimizer concedes transfer in an already
strong mode to lift weak target-relevant ones: it gives up the reference's tight
red concentration and draws the per-channel spot sizes toward a common value.
The change against optical proxies is from rewarding transferred contrast to
rewarding distinguishable target-relevant measurement modes, which is why a
control that raises MTF volume by \num{44.6}\% can still lose \num{17.4}\% of
the target information and image less well. Channel equalization is one possible
outcome of that trade rather than a requirement of it. The same width-only
optimization improves color on silicon-dioxide and titanium-dioxide libraries as
well, so the gain follows from the design criterion and not from one meta-atom
material.

This is a computational study of one camera architecture and fixed imaging
geometry, repeated across three meta-atom material libraries: the objective
depends on the scene prior, target, exposure and linear--Gaussian model, within which the target posterior and information are
exact. A learned nonlinear decoder under non-Gaussian statistics could favor
another allocation. The forward model is a local-periodic full-Jones RCWA chain,
independently verified against Tidy3D FDTD simulation of the complete aperture
(Results). Its numerical scope is stated in Supplementary Note~\snLimits. The
principle itself is not Bayer-specific: a different periodic sampler changes the
construction of $A_W$ and a different imaging task changes the target $T$.
Fabrication of the reference and information designs, which differ only in
their width map, together with paired tolerance analysis, are the immediate
next steps.

% =====================================================================
\section{Methods}

\subsection*{Full-Jones RCWA forward model}

The optic is a single layer of square silicon-nitride (Si$_3$N$_4$) pillars of
height \SI{1000}{nm} on fused silica, on a square lattice of pitch
\SI{290}{nm}, with widths confined to the usable interval. Both
materials are dispersive\citep{luke2015,malitson1965}. The circular pupil has
diameter $D=\SI{208}{\micro\meter}$ on a
\num{720}$\times$\num{720} square lattice and focal distance
$F=\SI{346.7}{\micro\meter}$. Thus $D/(2F)=0.3$ is the nominal paraxial NA,
and the exact geometric air-side value is
$\sin[\tan^{-1}(D/2F)]=\num{0.2873}$. The square grid contains
\num{518400} lattice sites, not active pillars. Active posts and independent
active variables are counted from the circular aperture mask.

The selected physical stack is a semi-infinite fused-silica source/object
half-space, followed by the SiN posts and air-side propagation to the sensor.
No finite entrance interface or finite substrate thickness is part of this
model, so no additional entrance-surface refraction is applied. Field
angles are substrate-side geometric angles, and conservation of tangential wave
vector across the patterned substrate-to-air stack is contained in the Jones
response and its input/output basis conversion.
The port media, polarization gauge and power normalization are recorded with the
response, and the angular-spectrum propagation uses the same air output medium.
The forward model uses the complete complex polarization response, square-cell C4
completion in azimuth, basis-consistent polarization transport and vectorial
propagation. Its local response is the complex
local-basis Jones matrix
\begin{equation}
J(W,\lambda,\theta,\phi)=
\begin{pmatrix}t_{ss}&t_{sp}\\t_{ps}&t_{pp}\end{pmatrix},
\label{eq:jones}
\end{equation}
tabulated in width, wavelength, polar angle and azimuth by rigorous coupled-wave
analysis\citep{moharam1981,moharam1995,li1996,kim2023torcwa}. Complex entries,
not standalone unwrapped phases, are interpolated. The azimuthal table is
completed from directly solved canonical azimuths by the exact C4
symmetry of the square post and square lattice. The table is solved at Fourier
order $(7,7)$ on a $256^2$ raster and C4-expanded from six directly solved
azimuths. The same table underlies every design reported here. The table's
high-order accuracy and the grid and C4 completion are discussed in Supplementary
Notes~\snLibrary{} and \snLimits, with the library and response-table
specification in Supplementary Table~1. The local-periodic approximation remains an
explicit modeling assumption\citep{pestourie2018}.

The stored matrix uses the solver's power-normalized modal amplitudes. The
zeroth transmitted order conserves tangential wave vector, so the input
silica-side and output air-side s/p bases are constructed separately, with
$n_{\rm in}\sin\theta_{\rm in}=n_{\rm out}\sin\theta_{\rm out}$.
Before Cartesian-field propagation the factor
$\sqrt{n_{\rm in}\cos\theta_{\rm in}/
(n_{\rm out}\cos\theta_{\rm out})}$ converts the modal Jones result to
electric-field amplitude. A non-propagating output order is excluded rather
than treated as an air-side ray at the input angle.

For each object point, the finite-conjugate incident phase is accumulated in
double precision before reduction modulo $2\pi$. The incident field is resolved
into local s and p components, multiplied coherently by all four entries of
Eq.~\eqref{eq:jones} and transformed back to the global transverse basis.
Non-paraxial angular-spectrum propagation follows the Fourier-optics
construction\citep{goodman2017}.
Orthogonal source polarizations are combined incoherently only after their
coherent field propagation. Pupil and detector grids share a matched physical
Fourier extent before inverse transformation and detector-plane cropping.

\subsection*{Sensor model and electron transfer}

At the sensor the transverse field passes a spatially invariant planar
back-illuminated dielectric stack, after which the
longitudinal and magnetic components are completed inside the silicon detection
medium from the divergence-free condition and Maxwell's curl equation, with the
magnetic field normalized by the vacuum wavenumber. The detected quantity is the
axial time-averaged Poynting component, summed incoherently over three
orthogonal dipole orientations for unpolarized light and over object points. The
resulting per-wavelength intensity is converted to normalized model-electron
units by a fixed exposure calibration and spectral quadrature over the measured
CFA transmissions and mean scene spectral-radiance spectrum. No periodic
microlens or iso-cell aperture is applied, and this planar unit-fill preserves
local shift invariance. The
\SI{1.0}{\micro\meter} photosite aperture is a sliding convolution evaluated
before sampling on the \SI{0.25}{\micro\meter} pre-sampling grid, and explicit
readout sampling then occurs inside the multirate operator of Supplementary
Note~\snOperator. The sliding $4\times4$ sum is multiplied by the fine-scene
sample-area factor
$(\SI{0.25}{\micro\meter}/\SI{1.0}{\micro\meter})^2=1/16$,
which preserves constant-radiance DC and the fixed one-micrometre electron
calibration. The optimization grid contains nine wavelengths from
\SIrange{420}{670}{nm}. Quantum efficiency and every omitted throughput
factor are stated explicitly in the electron calibration record. One exposure
scalar is set at a fixed reference anchor and then reused for every design
and field. Poisson shot noise enters the analytic objective through its
Gaussian-equivalent variance, with the read variance added in electron units.

\subsection*{Polyphase measurement operator}

The corrected sampling chain runs in one fixed order. Optical fields,
the scene and the sliding physical pixel-aperture convolution are all resolved
at \SI{0.25}{\micro\meter}. Convolution and pixel-aperture integration precede
sampling onto the \SI{1.0}{\micro\meter} readout lattice, and CFA multiplication and
two-by-two RGGB decimation follow. The resulting coarse RGGB lattice has
\SI{2.0}{\micro\meter} pitch. At the shortest wavelength, the exact geometric
air-side NA gives an incoherent cutoff of approximately
\SI{1.37}{\cycle\per\micro\meter}, so the \SI{0.25}{\micro\meter} grid
has a \SI{2.0}{\cycle\per\micro\meter} Nyquist limit, so the blue optical band
is fully represented before sampling. Unshifted two-by-two decimation onto the
RGGB lattice is periodic, so each coarse frequency observes \num{64} scene
aliases, and optical-to-pixel aliasing is formed before content is discarded. The
resulting $4\times(64K)$ measurement operator, together with the
$12\times(64K)$ target operator, is used throughout the optimization. The
derivation and self-conjugate-frequency treatment are given in Supplementary
Notes~\snOperator{} and \snPrior.

The prior is $\mathbf{C}(\nu)=S_x(\nu)C$, where $C$ is the dimensionless normalized
shape of the empirical nine-channel color covariance over natural
surfaces\citep{parkkinen1989,chakrabarti2011}, and $S_x$ is the spatial spectrum
carrying the sole variance scale in the normalized-radiance coordinate.
The normalization was verified by recovering the pixel-domain covariance
from the frequency-domain sum and by Monte Carlo sample covariance. The target
matrix $T$ maps the nine spectral-radiance channels to linear sRGB.

\subsection*{Design objective and optimization}

The objective is $\Itar$ of Eq.~\eqref{eq:targetinfo}. For each coarse frequency
and field, write $A_q\in\mathbb C^{4\times K}$ and
$T_q\in\mathbb C^{12\times K}$ for the measurement and target blocks of alias
$q$, and $\mathbf{C}_q\in\mathbb C^{K\times K}$ for its prior block. The exact
low-memory evaluation is
\begin{align}
\mathbf{C}_Y&=\mathbf{C}_{\mathrm n}+\sum_{q=1}^{64}A_q\mathbf{C}_qA_q^\dagger,
&
\mathbf{C}_Z&=\sum_{q=1}^{64}T_q\mathbf{C}_qT_q^\dagger,\nonumber\\
\mathbf{C}_{YZ}&=\sum_{q=1}^{64}A_q\mathbf{C}_qT_q^\dagger,
&
\mathbf{C}_{Y|Z}&=\mathbf{C}_Y-\mathbf{C}_{YZ}\mathbf{C}_Z^{-1}\mathbf{C}_{YZ}^\dagger,
\label{eq:measurement-space-loss}\\
\mathcal J_{\mathrm{tar}}(\nu,\theta)
&=\frac12\log_2\frac{\det\mathbf{C}_Y}{\det\mathbf{C}_{Y|Z}}.\nonumber
\end{align}
The frequency average, raw-pixel factor $1/4$ and field weights in
Eq.~\eqref{eq:targetinfo} then give $\Itar$. This form is algebraically
equivalent to the target-posterior determinant but streams the 64 independent
$K\times K$ prior blocks and factors matrices no larger than $12\times12$. The
implementation was verified by Cholesky solvability, positive-semidefiniteness
and a dense small-grid equivalence check.

The optimization machinery is implemented in PyTorch\citep{paszke2019} and uses
Adam\citep{kingma2015adam} under a cosine learning-rate schedule. Optimization
runs one stage of \num{300} mirror-quadrant steps at a learning rate of
\num{1.6e-2} decaying to \num{1e-5}, from a fixed hyperbolic reference seed
identical for every arm, with four stochastic field samples per step and
deterministic 25-field validation every ten steps.
Exact projected updates enforce the usable response-library interval. This
defines the currently implemented $\mathcal F$ in
Eq.~\eqref{eq:design-principle}. No smoothness or fabrication-barrier term is
added to $-\Itar$.

The best saved state is selected by deterministic full-quadrature evaluation and
then given a five-step all-field refinement at a constant rate of \num{1e-4} for
each arm. Stored full maps are rescored through the common evaluator. The stage
schedule and seeds are specified in Supplementary Note~\snOpt, with the fixed
protocol constants in Supplementary Table~3. For the
material-generality result the same objective, seed construction, schedule and
feasible-set projection were applied to the silicon-dioxide and titanium-dioxide
response libraries, with every other platform parameter unchanged.

\subsection*{Evaluation protocol}

Information, charge and primary imaging use a fixed common exposure. The
electron scale is fixed by the separate finite-conjugate \SI{540}{nm}
hyperbolic calibration anchor. The \SI{600}{nm} hyperbolic comparison reference,
MTF-volume design and information design then reuse that scale, so lower
throughput incurs lower signal and the corresponding shot noise.
Equal-brightness rendering is a secondary diagnostic and is never mixed with a
common-exposure claim.

$\Itar$ is the field-quadrature-weighted mean over the design grid at
the fixed exposure and empirical prior. Charge is reported both as mean
model-electron count per raw pixel, $(R+2G+B)/4$ for an RGGB cell, and by
channel. Field-resolved values are read from the saved 25-field evaluator on the
design quadrature.

Image formation and metric evaluation are separate steps. The forward simulator
writes raw model-electron, reconstruction, PSF, OTF and longitudinal/axial
diagnostic arrays with physical axes, and the evaluator computes metrics from
those saved arrays without rerunning the optical simulation.

Decoder hyperparameters are fixed independently, and a
fixed white-balance or color-correction transform is obtained per design from
a fixed analytic reference-camera DC mapping. Each design's transform is
analytic and fixed before it sees any test scene. No per-scene ground-truth fit
is used. The hyperbolic, MTF-volume and
information designs use identical reconstruction settings. Baseline construction
is described in Supplementary
Note~\snBaselines: both optimized arms use matched physical degrees of freedom,
width library, wavelength and field sampling, step budget and tuning protocol,
and native-objective convergence is verified before their stored maps are
evaluated by the common camera model.

\subsection*{Quantitative analysis}

The \num{80}\% encircled-energy radius is the radius of the circle, centered on
the photometric centroid of the point-spread function, containing \num{80}\% of
its energy, computed per wavelength on the fine grid and per color channel
after integration against the filter transmissions. PSNR is computed on linear
sRGB against the ground truth. The primary implemented color metric
is CIEDE2000 $\Delta E_{00}$ under a D65 white point, evaluated with the
standard unit parametric factors and checked against the published
Sharma--Wu--Dalal reference pairs\citep{sharma2005}. The older Euclidean
$\Delta E_{76}$ is retained only as a labeled diagnostic.
S-CIELAB applies opponent-channel spatial filtering before the color-distance
calculation\citep{zhang1997scielab}. These three metrics form the main imaging
summary. Metric definitions are given in Supplementary Note~\snImaging.

% =====================================================================
\section*{Data availability}

The point-spread-function cubes, width maps, target-information records,
raw model-electron arrays, rendered scenes and metric tables that support the
findings of this study are
being deposited in a public Zenodo archive. The digital object identifier will
be provided before publication, and the data are available from the author in
the interim.

\section*{Code availability}

The full-Jones RCWA forward model, target-information objective, constrained
optimizer and imaging pipeline used to design and score the computational
metalens are publicly available at
\url{https://github.com/hyoseokp/metalens-information-design}. The three
optimized width maps (hyperbolic reference, information design, MTF-volume
control) are included in the repository.

\section*{Author contributions}

H.P. conceived the study, developed the full-Jones RCWA forward model and
the objective, performed the optimization and the evaluation, prepared the
figures and wrote the manuscript. Y.P. contributed conceptualization,
supervision, and writing (review and editing), and is the corresponding author.

\section*{Competing interests}

The authors declare no competing interests.

\clearpage
\setcounter{figure}{0}
\renewcommand{\thefigure}{S\arabic{figure}}
\setcounter{table}{0}
\renewcommand{\thetable}{S\arabic{table}}
\setcounter{equation}{0}
\renewcommand{\theequation}{S\arabic{equation}}
\renewcommand{\figurename}{Supplementary Fig.}
\renewcommand{\tablename}{Supplementary Table}
\begin{center}
{\LARGE\bfseries Supplementary Information}
\end{center}

\noindent
This Supplementary Information provides the derivations, numerical methods,
optimization protocol and additional results supporting the main text. All
results are computational. All imaging evaluations are post-optimization checks
and do not enter the design objective. They cover one primary held-out scene and
seven additional held-out natural scenes (\S9.5).

\medskip\noindent
\textbf{Units and naming.} ``Electron'' and \eminus{} denote a calibrated
\emph{model-electron count}, the output of the normalized optical model after one
exposure calibration, and no number in this Supplement is an SI-radiometric
photoelectron prediction. The comparison reference is hyp600, the \SI{600}{nm}
hyperbolic width map. The corrected \SI{540}{nm} hyperbolic map (hyp540) sets the
exposure scalar, which is reused for hyp600 and every evaluated design.
Optimization is initialized from hyp600 projected onto the feasible set, and the
information design maximizes the field-averaged camera-delivered target
information
$\mathcal I_{\mathrm{tar}}=\sum_\theta w_\theta I(Y_{W,\theta};Z_\theta)$.

\medskip\noindent
\textbf{Contents.}
Supplementary Note~1, meta-atom library and RCWA response;
Note~2, the multirate RGGB polyphase operator;
Note~3, scene prior, aliased covariance and the self-conjugate-line edge case;
Note~4, camera-delivered target information, posterior entropy and its relation
to estimation;
Note~5, target-mode allocation and the restricted squared-transfer limit;
Note~6, optimization protocol and the measurement convention;
Note~7, MTF-volume baseline construction;
Note~8, imaging pipeline, decoder and metric definitions;
Note~9, extended results: information decomposition, axial focal structure,
wavelength-resolved transfer, imaging with the MTF-volume control and scene
robustness;
Note~10, stated limitations.

% =====================================================================
\snote{1: Meta-atom library and RCWA response}
\label{sn:library}

\subsection*{1.1 Geometry and materials}

The meta-atom is a square silicon-nitride post standing on a fused-silica
substrate, on a square lattice of pitch $P=290$\,nm, with a single fixed height
$h=1000$\,nm. The post width $W$ is the only geometric degree of freedom, and it
is confined throughout to $[100,240]$\,nm.

Dispersion is modeled by Sellmeier forms with $\lambda$ in micrometres. For the
silicon nitride~\citep{luke2015},
\begin{equation}
n_{\mathrm{SiN}}^2(\lambda)=1+\frac{3.0249\,\lambda^2}{\lambda^2-0.1353^2}
+\frac{40314\,\lambda^2}{\lambda^2-1239.8^2},
\label{eq:nsin}
\end{equation}
and for the fused-silica substrate the standard three-term form~\citep{malitson1965}
\begin{equation}
n_{\mathrm{SiO_2}}^2(\lambda)=1+\frac{0.6961663\,\lambda^2}{\lambda^2-0.0684043^2}
+\frac{0.4079426\,\lambda^2}{\lambda^2-0.1162414^2}
+\frac{0.8974794\,\lambda^2}{\lambda^2-9.896161^2}.
\label{eq:nsio2}
\end{equation}
Both media are treated as lossless and non-magnetic over the sampled band. The
superstrate is air.

\subsection*{1.2 The full-Jones RCWA response table}

The response of each meta-atom is the complete complex $2\times2$ Jones matrix
$J(W,\lambda,\theta,\phi)$ for substrate-side incidence and air-side output,
including both cross-polarized terms and physical-stack metadata, computed by
rigorous coupled-wave analysis~\citep{moharam1981,moharam1995,li1996} with a
GPU Fourier modal implementation~\citep{kim2023torcwa}. The post cross-section
is rasterized onto a $256\times256$ grid within the unit cell using an
anti-aliased midpoint coverage rule, so that the permittivity of a boundary cell
is weighted by the fraction of the cell the post covers. This removes the
staircase sensitivity that a binary raster would introduce into
$\partial\phi/\partial W$, which the optimizer differentiates through.
Supplementary Table~\ref{tab:library} collects the library specification.

\begin{table}[htbp]
\centering
\caption{\textbf{Meta-atom library and response-table specification.}}
\label{tab:library}
\small
\begin{tabular}{ll}
\toprule
Quantity & Value\\
\midrule
Post material / cross-section      & silicon nitride, square\\
Substrate / superstrate            & fused silica / air\\
SiN index model                    & Sellmeier, Eq.~\eqref{eq:nsin}\\
Substrate index model              & Sellmeier, Eq.~\eqref{eq:nsio2}\\
Lattice                            & square, pitch $P=290$\,nm\\
Post height                        & $h=1000$\,nm (fixed)\\
Post width range used by designs   & 100--240\,nm (hard bound)\\
\midrule
Solver                             & rigorous coupled-wave analysis, GPU Fourier modal method\\
Fourier truncation order           & 7 per transverse direction\\
Unit-cell raster                   & $256\times256$, anti-aliased midpoint coverage\\
Tabulated quantity                 & complex $2\times2$ Jones matrix, zeroth transmitted order\\
Orientation                        & substrate-side incidence, air-side output\\
\midrule
Width grid                         & 100--240\,nm, 2.5\,nm step\\
Wavelength grid                    & the nine wavelengths (Note~2)\\
Polar-angle grid                   & 0--24$^\circ$, 1$^\circ$ step (substrate side)\\
Azimuth grid                       & $-180^\circ$--$180^\circ$, 15$^\circ$ step, C4-completed\\
\midrule
Mean power transmittance $|t|^2$   & 0.947 over the design window\\
Median power transmittance         & 0.969\\
Fraction of cells with $|t|^2<0.8$ & 4.2\%\\
Unwrapped phase span across width  & 5.59--10.44\,rad, depending on wavelength\\
\bottomrule
\end{tabular}
\end{table}

Complex Jones matrices are directly solved at canonical azimuths 0--75$^\circ$
and the remaining azimuths are completed by the C4 symmetry of the square post
and square lattice, with no reciprocity or other point-group operation applied
implicitly.

The library is efficient without being lossless. Over the design window
(post widths 100--240\,nm, wavelengths 420--670\,nm) the mean power transmittance
is 0.947 and the median 0.969, but the distribution has a low tail: 4.2\% of
(width, wavelength) cells fall below $|t|^2=0.8$ and 1.0\% below 0.5, near a
guided-mode resonance of the post. Because the objective reflects the calibrated
model-electron response and not a DC-normalized transfer (\S2.3), a width map that parks
pillars in that tail is penalized automatically through $\mu_a(W)$, without any
efficiency term being added by hand.

The accessible phase spans approximately one $2\pi$ branch. The
unwrapped phase swept by the full width interval ranges from 5.59\,rad at the
short-wavelength end to 10.44\,rad at the long-wavelength end. At the shortest
wavelength the span falls slightly below a full $2\pi$, while longer wavelengths
provide more than one full cycle but never two. The
width-to-phase inversion used to construct the hyperbolic prescription
is for that reason restricted to a single monotone branch, and the
per-wavelength branch assignment is a discrete choice a gradient cannot cross
(\S6.4).

Within each material platform, the response is a function of $W$ alone at fixed $\lambda$ and $\theta$.

\subsection*{1.3 Diffraction orders and the Rayleigh threshold}

The tabulated quantity is the zeroth transmitted order alone, and a non-zero
order becomes propagating in the substrate over part of the design band. A
non-zero transmitted order $(m,n)$ propagates in the substrate when
\begin{equation}
\Big(n_i\sin\theta\cos\phi+\frac{m\lambda}{P}\Big)^{2}
+\Big(n_i\sin\theta\sin\phi+\frac{n\lambda}{P}\Big)^{2}<n_t(\lambda)^{2},
\label{eq:rayleigh}
\end{equation}
a two-dimensional condition on the square lattice with $n_t$ the dispersive silica
index and $\phi$ the incidence azimuth. Evaluated with the dispersive
$n_t(\lambda)$, the threshold wavelength is 425.6\,nm at normal incidence and
reaches a maximum of 499.4\,nm over the design quadrature. Of the nine design
wavelengths, 420 and 450\,nm lie above threshold at every field, 470\,nm over
60\% of the quadrature weight, and 510\,nm and every longer wavelength lie below
threshold everywhere.

The coupled-wave amplitudes are flux-normalized, so $|t|^2$ is the fraction of
incident power in order zero and the diffracted power is excluded from the
modeled zeroth-order field. Non-zero diffraction orders and their propagation in
a finite substrate are outside the present model.

\subsection*{1.4 Additional material platforms (generality study)}

The generality comparison (main-text Figs.~3 and~4) repeats the full optimization and imaging
pipeline with two additional meta-atom material stacks, using independently
built RCWA look-up tables.
Supplementary Table~\ref{tab:materials} collects the parameters that differ
across platforms. The optimization recipe (300 mirror-quadrant steps, cosine
learning-rate schedule $1.6\times10^{-2}$ to $10^{-5}$, field-sampling seed
20260805) and all sensor, pitch and width-bound parameters are identical for
every material.

\begin{table}[htbp]
\centering
\caption{\textbf{Per-material library specifications for the generality study.}
All three platforms share the square lattice ($P=290$~nm), width range
100--240~nm and substrate refractive index (fused silica,
Eq.~\eqref{eq:nsio2}). $^{\dagger}$cubic-interpolated to the 57-point grid
(see text).}
\label{tab:materials}
\small
\begin{tabular}{lccc}
\toprule
 & SiN & SiO$_2$ & TiO$_2$\\
\midrule
Post material & silicon nitride & fused silica & amorphous TiO$_2$\\
Post height & 1000\,nm & 2500\,nm & 700\,nm\\
Index model & Eq.~\eqref{eq:nsin} & Eq.~\eqref{eq:nsio2} & Siefke 2016 (2-term)~\citep{siefke2016}\\
Superstrate & air & air & air\\
\midrule
RCWA truncation order & 7 & 5 & 7\\
Unit-cell raster & $256^2$ & $128^2$ & $256^2$\\
Width grid & 57 pts (2.5\,nm step) & 21 pts (7\,nm step)$^{\dagger}$ & 57 pts (2.5\,nm step)\\
\midrule
$\Itar$ (bit/raw px) & 0.5860 & 0.5776 & 0.5984\\
\bottomrule
\end{tabular}
\end{table}

The SiO$_2$ library uses a coarser solver grid (order~5, raster $128^2$, 21
width points) and is cubic-interpolated to the standard 57-point grid. The
low refractive index of the all-glass stack produces a smooth, resonance-free
width-to-phase response, so the interpolated table reproduces the directly
solved points to better than $10^{-3}$\,rad phase error.

% =====================================================================
\snote{2: The multirate RGGB polyphase operator}
\label{sn:operator}

A color camera records one scalar model-electron count per pixel, drawn from a
color-filter array (CFA) that repeats with period two in each direction. It
records neither wavelength-resolved point-spread functions (PSFs) nor three
continuously sampled RGB images. This note derives the exact linear operator that
maps a locally stationary spectral scene to that raw frame. Everything downstream
(the posterior, the target-information objective and its design gradient) is
then computed on the measurement the camera makes rather than on a proxy for it.

\subsection*{2.1 Geometry, sampling and conventions}

The design variable is the pillar-width map $W(x,y)$ on a $720\times720$ square
lattice of pitch 0.29\,\um. The optical aperture $A(x,y)$ is the inscribed disc
of diameter $D=208$\,\um{} (the lattice itself spans 208.8\,\um), and the focal
length is $f=346.7$\,\um. Thus $D/(2f)=0.3$ is the nominal paraxial NA,
whereas the exact air-side geometric value is
$\sin[\tan^{-1}(D/(2f))]=0.2873$. Pillars are
square silicon-nitride posts of a single fixed height of 1.0\,\um{} on a 290\,nm
pitch, with widths confined to the fixed, library-supported interval
$W\in[0.10,0.24]$\,\um. The object plane sits at 47{,}697\,\um. Production
optical intensities and scene spectra are represented on an $832\times832$ grid
of pitch 0.25\,\um. The 1.0\,\um{} square photosite aperture is applied as a
\emph{sliding presampling convolution} on that fine grid. Only afterwards does
the raw lattice select every fourth fine sample, giving $208\times208$ samples
at 1.0\,\um{} pitch. CFA multiplication and two-by-two phase extraction then
give the $104\times104$ RGGB-cell grid.

The 0.25\,\um{} scene grid has a 2\,cyc/\um{} Nyquist limit, above the
approximately 1.37\,cyc/\um{} blue incoherent cutoff of the exact NA,
so the blue optical band is fully represented before sampling.
Production Fourier states use an unshifted \emph{unitary} DFT with DC at
$[0,0]$, and every paired power spectrum, mask, signal covariance and noise
covariance must use that same normalization.
Supplementary Table~\ref{tab:constants} collects the protocol constants that are
fixed across every design, evaluation and figure in this work.

\begin{table}[htbp]
\centering
\caption{\textbf{Production protocol constants.}}
\label{tab:constants}
\footnotesize
\begin{tabular}{@{}p{0.10\linewidth}p{0.31\linewidth}p{0.54\linewidth}@{}}
\toprule
Group & Quantity & Value\\
\midrule
Optic      & aperture diameter $D$                & 208\,\um\\
           & nominal paraxial / exact air NA      & 0.30 / 0.2873\\
           & focal length $f$                     & 346.7\,\um\\
           & pillar material / cross-section      & silicon nitride, square post\\
           & pillar lattice pitch                 & 290\,nm\\
           & pillar height                        & 1.0\,\um{} (single fixed value)\\
           & width interval (hard bound)          & 0.10--0.24\,\um\\
           & square-grid lattice sites            & $720\times720=518{,}400$\\
           & active posts inside circular aperture& 404{,}048 (geometry count)\\
           & active independent sites (one mirror quadrant) & 101{,}012 (the independent optimization variables)\\
           & camera orientation                   & substrate incidence $\rightarrow$ posts $\rightarrow$ air gap / sensor\\
           & response table                       & substrate-to-air full-Jones RCWA tabulation, Fourier order $(7,7)$, C4 completion (\S1.2)\\
           & propagation medium index             & 1 (air gap beyond the pillars)\\
\midrule
Sampling   & object distance                      & 47{,}697\,\um\\
           & propagation pad factor               & 2 (transform grid doubled per axis)\\
           & pupil grid                           & $720\times720$\\
           & scene / presampling grid  & $832\times832$ / 0.25\,\um\\
           & photosite aperture                   & 1.0\,\um{} square, sliding fine-grid convolution\\
           & readout pixel grid / pitch           & $208\times208$ / 1.0\,\um\\
           & coarse Bayer grid                    & $104\times104$\\
           & optimization wavelengths (9)         & 420, 450, 470, 510, 540, 570, 600, 635, 670\,nm\\
\midrule
Sensor     & color filter array                  & RGGB\\
           & detector spatial model            & planar BSI transfer, unit fill; no periodic microlens/iso-cell\\
           & calibration anchor                   & corrected medium-aware hyp540: 1000 model-\eminus; hyp600 evaluated at the same fixed exposure\\
           & read noise                           & 1.5\,\eminus{} rms\\
           & model-electron calibration scalar $\kappa_{\mathrm{cal}}$ & fixed scalar $8.4112\times10^{-13}$\\
\midrule
Scene prior& latent dimension $K$                 & 9 (empirical normalized-radiance covariance shape)\\
           & spatial PSD                          & $\alpha_{\mathrm{psd}}/(k_0+|f|)^{\beta}$\\
           & $k_0$, $\beta$                       & 0.02\,cyc/\um{} (0.005\,cyc/fine sample), 2\\
           & contrast rms (normalized-radiance units) & 0.18\\
           & reconstruction target                & linear sRGB, $T\in\mathbb R^{3\times9}$\\
\midrule
Field      & quadrature                           & 25 fields, square first quadrant, Snell/sensor/Nyquist gated\\
\bottomrule
\end{tabular}
\end{table}

\subsection*{2.2 Meta-atom response and the per-wavelength pixel PSF}

The forward model applies the complete complex power-normalized Jones operator
\begin{equation}
\mathbf a^{(P)}_{\mathrm{out}}=
J_P(W,\lambda,\theta_{\rm in},\phi)\,
\mathbf a^{(P)}_{\mathrm{in}},
\label{eq:si-jones}
\end{equation}
with angle-dependent co-polarized amplitude and phase, both cross-polarized
terms and the substrate-to-air orientation. Here the superscript $(P)$
denotes the solver convention for which squared modal amplitude is a power
ratio. Tangential wave vector is conserved across the zeroth transmitted order,
\begin{equation}
n_{\rm in}\sin\theta_{\rm in}=n_{\rm out}\sin\theta_{\rm out},\qquad
\mathbf a^{(P)}_{\rm in}=\sqrt{n_{\rm in}\cos\theta_{\rm in}}\,
\mathbf a^{(E)}_{\rm in},\qquad
\mathbf a^{(E)}_{\rm out}=
\frac{\mathbf a^{(P)}_{\rm out}}
{\sqrt{n_{\rm out}\cos\theta_{\rm out}}}.
\label{eq:jones-power-to-e}
\end{equation}
Thus the incident silica-side and transmitted air-side p axes are constructed
from different wave vectors even though their s axes share an azimuth. The
normal-incidence electric-field factor is already
$\sqrt{n_{\rm in}/n_{\rm out}}$, so treating $J_P$ as a raw electric-field matrix
would be wrong. A non-propagating output order is rejected. Treating
each post by the response of an infinite array of identical posts is the
local-periodic approximation~\citep{pestourie2018}.

The remaining forward optics comprise the spherical incident field, aperture and
meta-atom modulation, exact non-paraxial vectorial angular-spectrum propagation
over the pupil--sensor gap, the back-illuminated pixel stack, and completion of
the longitudinal and magnetic field components in silicon followed by axial
Poynting detection. For a single object point at field angle $\theta$ they
produce the incoherent per-wavelength intensity PSFs
$\mathrm{PSF}_l(\mathbf r;\theta)$ at the $L=9$ sampled wavelengths
\begin{equation}
\lambda\in\{420,450,470,510,540,570,600,635,670\}\ \mathrm{nm}.
\label{eq:wl}
\end{equation}
Let $a_{\rm pix}$ be the sampled 1.0\,\um{} square photosite aperture on the
0.25\,\um{} grid. The presampling response is a sliding convolution
on the padded fine grid
\begin{equation}
p_l^{\rm fine}(\mathbf r;\theta)=
\left(\frac{\Delta_{\rm fine}}{\Delta_{\rm pix}}\right)^2
\big(a_{\rm pix}*\mathrm{PSF}_l\big)(\mathbf r;\theta),
\qquad
\left(\frac{\Delta_{\rm fine}}{\Delta_{\rm pix}}\right)^2
=\left(\frac{0.25}{1.0}\right)^2=\frac1{16},
\qquad \mathbf r\in\{0,\dots,831\}^2,
\label{eq:pixelpsf}
\end{equation}
where $a_{\rm pix}$ is the unnormalized $4\times4$ support. The explicit
fine-scene sample-area factor preserves constant-radiance DC and the fixed
1\,\um{} electron calibration. A bare sliding sum would spuriously multiply
the charge and shot-noise operating point by 16 when the latent grid is refined.
Its fine-grid Fourier transform seeds $M^{\rm fine}$ in
Eq.~\eqref{eq:operator-prod}, and lattice samples are not selected until that
operator is assembled.

\medskip\noindent\emph{Boundary conditions.} The selected
camera model has a semi-infinite fused-silica source/object half-space, followed
by the posts, an air gap and the sensor. Propagation over the 346.7\,\um{}
post-to-sensor gap therefore uses the air dispersion relation, whereas the
object-to-pupil wave vector uses $n_{\mathrm{SiO_2}}(\lambda)k_0$. All LUT and
field angles are substrate-side geometric angles. No finite entrance interface,
finite substrate thickness or Snell conversion is part of this model.
The full-Jones RCWA LUT is consequently solved for fused-silica incidence and
air output and records the angle definition, power normalization, polarization
gauge and port media explicitly.

\subsection*{2.3 Calibrated common-exposure model-electron transfer}

For sensor channel $a\in\{R,G,B\}$ and wavelength $\lambda_l$, the normalized
model-electron weight combines the trapezoid spectral quadrature, the
photon-per-energy factor, and the color-filter and placeholder
quantum-efficiency response,
\begin{equation}
w^{(e)}_{a,l}=r_a(\lambda_l)\,s(\lambda_l)\,\Delta\lambda_l\,\frac{\lambda_l}{hc},
\qquad r_a(\lambda_l)=\mathrm{CFA}_a(\lambda_l)\,\mathrm{QE}(\lambda_l),
\label{eq:eweight}
\end{equation}
where $r_a(\lambda_l)$ is the color-filter times quantum-efficiency response,
$s(\lambda_l)$ the mean normalized scene spectrum, $\Delta\lambda_l$ the
trapezoid quadrature weight and $\lambda_l/hc$ the photon-per-energy factor,
with $h$ Planck's constant, $c$ the speed of light, $\mathrm{QE}=1$ at every
sampled wavelength, and $\Delta\lambda_l$ the
irregular-grid trapezoid weight in metres
($\Delta\lambda_0=\tfrac12(\lambda_1-\lambda_0)$,
$\Delta\lambda_l=\tfrac12(\lambda_{l+1}-\lambda_{l-1})$ in the interior,
$\Delta\lambda_{L-1}=\tfrac12(\lambda_{L-1}-\lambda_{L-2})$). Here
$s(\lambda_l)$ is the \emph{mean normalized scene spectrum} of the empirical
prior: the weighted mean, over the Cartesian product of the measured surface
reflectances of the dataset with its daylight illuminants (each normalized to
unit luminance), in the dataset's normalized-radiance coordinate. It is not a
physical radiance in SI units. Its dimensionless luminance is $Y=0.1666$ under
that dataset normalization. The complex calibrated model-electron optical
transfer function is
\begin{equation}
\mathrm{OTF}^{(e)}_{a,l}(\nu;\theta)=\kappa_{\mathrm{cal}}\,w^{(e)}_{a,l}\,
\widehat{p_l}(\nu;\theta),\qquad \widehat{p_l}=\mathrm{fft2}[p_l].
\label{eq:otf}
\end{equation}
The scalar $\rho$ is the dimensionless normalization of the flat-spectrum point
probe used to generate $p_l$. It factors out of the wavelength sum and is fixed
by the PSF protocol. The fixed $\kappa_{\mathrm{cal}}$ maps that normalized
response to the calibrated model-electron scale. The channel mean model-electron
count is then the DC response,
\begin{equation}
\mu_a(W;\theta)=\mathrm{Re}\Big[\textstyle\sum_l\mathrm{OTF}^{(e)}_{a,l}(0;\theta)\Big]
=\kappa_{\mathrm{cal}}\,\rho\,g_a(W;\theta),
\label{eq:mean}
\end{equation}
which factorizes into the fixed calibration scalar, the dimensionless probe
normalization $\rho$, and the design- and channel-dependent throughput
$g_a(W;\theta)=\sum_l r_a(\lambda_l)s_l\Delta\lambda_l(\lambda_l/hc)\,
\tilde p_l(0;\theta)$ with $\tilde p_l=p_l/\rho$. Equivalently $\mu=M(0)\,s$:
the mean model-electron count is produced by the normalized prior mean.
That identity makes the linearization below consistent, because the shot-noise
variance is evaluated at the mean of the same latent state mapped by the camera
operator.

\medskip\noindent\emph{Linearization about a mean scene.} The scene state that
enters the polyphase construction is zero-mean,
$x\sim\mathcal{CN}(0,\mathbf{C}_{\mathrm{al}})$, whereas
$\mu_a$ in Eq.~\eqref{eq:mean} is a \emph{mean} model-electron count. The two are
reconciled by linearizing the readout about a mean scene. The signal is carried
by the zero-mean spatial contrast about that mean, while the shot-noise variance
is set by the normalized mean operating point and is therefore treated as
signal-independent within a tile. With that expansion the Gaussian
shot-plus-read noise covariance is diagonal in the sensor channels,
\begin{equation}
\sigma_a^2=\mu_a+\sigma_r^2,\qquad \sigma_r=1.5\ e^-_{\mathrm{model}}.
\label{eq:noise}
\end{equation}
As usual for a Poisson count, the numerical identity
$\operatorname{Var}N=\operatorname{E}N$ treats the count number as dimensionless.
We annotate the mean as model-\eminus{} and the resulting covariance as squared
model-\eminus.  This bookkeeping does not turn the normalized model into an
absolute radiometric calibration.
Equation~\eqref{eq:noise} therefore carries two approximations: a Gaussian
surrogate for Poisson shot noise, and a first-order expansion of the readout
about a mean scene. The design-dependence of $\mathbf{C}_{\mathrm n}$ through
$\mu_a(W)$, one of the properties the camera operator is intended to carry,
comes from that linearization. Section~2.6 lists it again in its scope statement.

\medskip\noindent\emph{The fixed model-electron calibration.} The 
scalar $\kappa_{\mathrm{cal}}$ is generated from the full-Jones RCWA LUT and
forward code. The anchor is a corrected finite-conjugate hyperbolic width map
built by inverting the same LUT at 540\,nm. Its required phase cancels
$k_0[n_{\rm in}(\lambda)\,\mathrm{OPD}_{\rm obj}
+n_{\rm out}(\lambda)\,\mathrm{OPD}_{\rm sen}]$. The scalar maps the brightest
channel of that anchor under the normalized mean scene spectrum of
Eq.~\eqref{eq:eweight} to 1000 model-\eminus{} and is then reused unchanged
for hyp600, every evaluated design and every field. Choosing another anchor
would define a different operating exposure and require a complete rescore.
The distinction is between a \emph{common-exposure} and a
\emph{fixed-signal} comparison. Under common exposure the operating means follow
the throughput, $\mu_a(W)=\kappa_{\mathrm{cal}}\rho\,g_a(W)$, so a design that
has lower normalized response in a channel is penalized with a lower $\mu_a$ and
a worse shot-limited signal-to-noise ratio. Recalibrating per design would impose
$\mu_a(W)\equiv\mu_{\mathrm{target}}$ through a design-dependent row gain
$c_a(W)=\mu_{\mathrm{target}}/(\kappa_{\mathrm{cal}}\rho\,g_a(W))$, which
\emph{cancels} $g_a(W)$ and erases exactly the throughput differences this study
measures. The optimizer uses the calibrated common-exposure transfer. Row-wise
and global fixed-signal transfers are shape-only diagnostics.

\medskip\noindent\emph{Imaging convention.} The primary whole-scene evaluation
of Supplementary Note~8 uses this same common exposure. Per-design
equal-brightness rendering is a secondary blur and allocation diagnostic and
does not feed a primary metric or figure.

\subsection*{2.4 Color-mixing transfer}

Let $x(\nu)\in\mathbb C^{K}$ be the latent scene state: the normalized spectral
radiance at the $K=L=9$ sampled wavelengths, in the model coordinate,
with prior mean $s$ and zero-mean fluctuation. The basis
$B\in\mathbb R^{K\times L}$ maps that normalized-radiance state into the
mean-normalized coordinates that the $s$-weighted transfer of
Eq.~\eqref{eq:eweight} consumes. For $K=L$ it is the fractional identity
$B=\operatorname{diag}(1/s_l)$, so $\operatorname{diag}(s)B=I_L$. Contracting
the per-wavelength transfer against it gives the model-electron color-mixing
transfer
\begin{equation}
M_{a,k}(\nu;\theta)=\sum_{l=1}^{L}\mathrm{OTF}^{(e)}_{a,l}(\nu;\theta)\,B_{k,l},
\qquad M(\nu)\in\mathbb C^{3\times K}.
\label{eq:mix}
\end{equation}
Because $B$ removes the $s$ that Eq.~\eqref{eq:eweight} inserted, $M$ is a
transfer \emph{per unit normalized spectral radiance} and carries no illuminant
factor of its own. The illuminant enters the model exactly once, as the
operating point about which the readout is linearized, and therefore appears in
$\mu_a$, in $\mathbf{C}_{\mathrm n}$ and in the prior mean, while correctly leaving
$M$ alone. In particular $M(0)s=\mu$.

$M$ retains the full complex, field-dependent phase and the calibrated
model-electron gain. $M$ is the matrix-valued object that replaces a scalar
per-channel transfer function. Supplementary Note~5 sets out the consequences of
that replacement.

\subsection*{2.5 The sixty-four-alias multirate operator}

The scene grid has spacing 0.25\,\um, while one RGGB cell has pitch 2\,\um.
The scene-to-cell decimation is therefore eight per axis and folds sixty-four
fine-grid aliases.  Let the fine scene grid have shape $H\times W$, let
\[
q=(q_y,q_x)\in\{0,\ldots,7\}^2,\qquad
r_p\in\{(0,0),(0,4),(4,0),(4,4)\}
\]
be, respectively, the alias index and the four RGGB phase offsets measured in
0.25\,\um{} scene samples, with the RGGB channel map
$c(p):\ (0,0)\mapsto R,\ (0,4)\mapsto G,\ (4,0)\mapsto G,\ (4,4)\mapsto B$.
The information model uses unitary fine- and coarse-grid DFTs, consistently
with the prior normalization of \S3.1. Define
\[
\omega_q(\nu)=2\pi\left(
\frac{\nu_y+q_yH/8}{H},
\frac{\nu_x+q_xW/8}{W}\right).
\]
If $M^{\rm fine}_{a,k}$ already contains optical propagation and the complete
physical photosite-aperture response \emph{before} sampling, the operator is
\begin{equation}
\boxed{\;
\big[A_W^{\rm prod}(\nu)\big]_{p;(q,k)}
=\frac1{8}\exp\!\big(i\,\omega_q(\nu)\!\cdot r_p\big)\,
M^{\rm fine}_{c(p),k}\!\left(
\nu+\left(q_y\tfrac H8,q_x\tfrac W8\right);\theta\right)\;}
\label{eq:operator-prod}
\end{equation}
and has shape $4\times64K$.  The desired image is a full-RGB image on the
1\,\um{} target-pixel grid, not a 0.25\,\um{} super-resolved target. Let
$h_{\rm tar}(\xi)$ be the 1\,\um{} square target-pixel aperture transfer. For target phase
$r\in\{(0,0),(0,4),(4,0),(4,4)\}$ and color row $c$, the corresponding
$12\times64K$ target operator is
\begin{equation}
\big[T_a^{\rm prod}(\nu)\big]_{(r,c);(q,k)}
=\frac1{8}\exp\!\big(i\,\omega_q(\nu)\!\cdot r\big)\,
h_{\rm tar}\!\left(\nu+\left(q_y\tfrac H8,q_x\tfrac W8\right)\right)T_{c,k}.
\label{eq:target-prod}
\end{equation}
Thus scene content above the 1\,\um{} target Nyquist limit can alias into the
measurement while the target remains the 1\,\um{} color image. The
$1/8$ coefficient is the two-dimensional unitary-DFT decimation coefficient:
$1/\sqrt8$ per axis.

The unitary convention applies to the scene coefficients on both grids.  It
does not mean that a spatial PSF is inserted as its unscaled unitary FFT.  If
$h$ is a kernel on the $H\times W$ fine grid, its convolution multiplier is
\begin{equation}
H_{\rm conv}=\sqrt{HW}\,\mathcal F_{\rm u}\{h\},
\label{eq:unitary-convolution-bridge}
\end{equation}
which is numerically the default backward-FFT transform of $h$.

The operator of Eq.~\eqref{eq:operator-prod} carries (i)~the calibrated
model-electron gain, through $M$, (ii)~the alias fold, through the sum over $q$,
(iii)~the CFA spectral crosstalk, through $c(p)$ and $r_a(\lambda)$, and
(iv)~the design-dependent noise, through
$\mathbf{C}_{\mathrm n}=\operatorname{diag}(\mu_a(W)+\sigma_r^2)$, the green
variance repeated for the two green phases. None of the four is visible to a
per-channel, DC-normalized scalar transfer functional (Supplementary Note~5).

\subsection*{2.6 Model assumptions}

Equation~\eqref{eq:operator-prod} is exact for a locally shift-invariant Gaussian
scene observed through an infinite periodic $2\times2$ mosaic, or equivalently
through its circular finite-grid analogue, \emph{given} the inputs it is built
from. Three qualifications limit how broadly ``exact'' should be read.

\begin{enumerate}
\item[(i)] \emph{Shift-invariance and periodicity.} The operator is not a global
raw operator for a spatially varying or finite-boundary optical system. Field
variation is handled \emph{outside} it, by the 25-point quadrature of
Supplementary Note~6 rather than inside it. This is a stated scope, not an approximation error. A field-integrated
number is a weighted sum of these local operators.
\item[(ii)] \emph{The noise input is linearized.} The noise covariance
comes from a first-order expansion of the readout about a mean
scene (\S2.3), so the design-dependence of $\mathbf{C}_{\mathrm n}$ through
$\mu_a(W)$ inherits that expansion, as well as the Poisson-to-Gaussian
surrogate.
\item[(iii)] \emph{The meta-atom response is local-periodic.} Each post is
assigned the response of an infinite array of identical posts (Supplementary
Note~1), tabulated at discrete angles. This is an explicit modeling
assumption whose error is not quantified for the map and stack
(Note~10).
\end{enumerate}

% =====================================================================
\snote{3: Scene prior, aliased covariance, and the self-conjugate-line edge case}
\label{sn:prior}

\subsection*{3.1 Stationary Gaussian scene with an isotropic power spectrum}

The scene is modeled as a stationary zero-mean Gaussian field with isotropic
power spectrum on the unshifted pixel-frequency grid,
\begin{equation}
S_x(f)=\frac{\alpha_{\mathrm{psd}}}{(k_0+|f|)^{\beta}},\qquad
k_0=0.02\ \mathrm{cyc/\um}=0.005\ \text{cyc/fine sample},\quad
\beta=2,
\label{eq:psd}
\end{equation}
with $\alpha_{\mathrm{psd}}$ fixed by normalizing the discrete full-grid mean to the contrast
variance,
\begin{equation}
\frac{1}{HW}\sum_f S_x(f)=\mathrm{contrast\_rms}^2=0.18^2=0.0324,
\label{eq:psdnorm}
\end{equation}
so that under the unitary-DFT covariance convention $S_x$ is the zero-mean
spatial-contrast variance density. That rms is the sole
normalized-radiance fluctuation scale after the dimensionless color-covariance
normalization of \S3.2, not a second scale carried by $C$ and not a
fractional contrast.
Measured against the prior's own mean scene luminance $Y=0.1666$, it is of the
same order as the mean itself, which is what building a prior over daylight
variation as well as over surfaces implies. The same figure is applied to every
design compared here. The $1/f^{\beta}$ form with $\beta\approx2$ is
the standard summary of natural-image spatial statistics~\citep{field1987,torralba2003}. The constant $k_0$
regularizes the DC divergence.

\subsection*{3.2 Latent color covariance and the reconstruction target}

The full latent spectral covariance is the product of the normalized-radiance
spatial PSD with a dimensionless $K\times K$ latent color-covariance shape $C$,
\begin{equation}
\mathbf{C}(\nu)=S_x(\nu)\,C\in\mathbb C^{K\times K}.
\label{eq:sigma}
\end{equation}
The prior used throughout is a fixed empirical prior with $K=L=9$. The raw
spectral covariance is estimated from the Cartesian product of measured surface
reflectances and daylight illuminants, then normalized so that
$\operatorname{tr}(T_{\mathrm{XYZ}}CT_{\mathrm{XYZ}}^\top)/3=1$, so $C$ is
dimensionless. The spatial PSD $S_x$ supplies the only contrast scale,
with frequency mean $0.18^2$ in squared normalized-radiance units. The dataset
mean spectrum is $s$. The spectral basis is the fractional identity
$B=\operatorname{diag}(1/s_l)$, so $\operatorname{diag}(s)B=I_9$ and each latent
coordinate is the normalized scene spectral radiance at one sampled wavelength,
with prior mean $s_l$. The reconstruction target $T\in\mathbb R^{3\times9}$ is
the CIE 1931 quadrature followed by the XYZ-to-linear-sRGB matrix, so it maps
that normalized-radiance state to linear sRGB. Low-dimensional linear models of
surface reflectance of exactly this kind are long
established~\citep{maloney1986,parkkinen1989}, and the second-order statistics
of natural hyperspectral scenes are similarly well
characterized~\citep{chakrabarti2011}. With this convention $\mathbf{C}$ has
normalized-radiance-squared units and $A_W\mathbf{C} A_W^\dagger$ has
model-electron-squared units, as required for addition to the noise covariance.
Because it is a single empirical prior, its scope is stated in \S3.4.

\subsection*{3.3 Independence of the aliases and self-conjugate coarse cells}

Because the scene is stationary and the grid periodic, distinct fine-grid
frequencies that are not related by conjugation are uncorrelated and, under the
Gaussian model, independent.  Away from the self-conjugate cells discussed
below, the sixty-four aliases of Eq.~\eqref{eq:operator-prod} therefore give
\begin{equation}
\mathbf{C}_{\mathrm{al}}^{\rm prod}(\nu)=
\bigoplus_{q\in\{0,\ldots,7\}^2}
\mathbf{C}\!\left(\nu+\left(q_y\tfrac H8,q_x\tfrac W8\right)\right)
\in\mathbb C^{64K\times64K}.
\label{eq:sigmaal}
\end{equation}

A real-valued scene satisfies $X[-k]=X^*[k]$. For a fine scene-grid length
$H=832$ and scene-to-RGGB-cell decimation eight, the coarse length is
$H/8=104$. A conjugate alias falls in the same coarse cell only when
\begin{equation}
u\equiv-u\pmod{104},\qquad u\in\{0,52\}.
\label{eq:selfconj}
\end{equation}
The corresponding two-dimensional set contains four coarse cells:
$(0,0)$, $(0,52)$, $(52,0)$ and $(52,52)$. Within those cells some of the
sixty-four aliases are conjugate pairs and some coefficients are self-conjugate.
The circular-complex block construction is exact there for every Hermitian
prior, so it needs no correction in general. The implemented real-field
construction instead separates real and imaginary coordinates and drops their
cross-covariance, which is exact precisely when the prior covariance is real.
The prior used here is real, so the two constructions agree numerically. A
complex-valued prior would make the real-field path an approximation rather than
an identity.
The Q64 core converts each conjugate pair to independent real and imaginary
coordinates with covariance $C/2$, retains a real singleton with covariance
$C$.

\subsection*{3.4 Dependence on the scene prior}

Every target-information value in this manuscript is computed under the one
empirical prior specified above, which enters the objective through
$\mathbf{C}_{\mathrm{al}}$. The ranking is a statement about scenes drawn from this
prior.

% =====================================================================
\snote{4: Camera-delivered target information}
\label{sn:info}

\subsection*{4.1 Target and measurement}

At each field point and coarse spatial frequency, the physical camera is written
as
\begin{equation}
Y_W(\nu,\theta)=A_W(\nu,\theta)X(\nu)+N(\nu,\theta),
\qquad Z(\nu)=T_aX(\nu),
\label{eq:camera-target}
\end{equation}
where $A_W$ contains wavelength-dependent propagation, calibrated
common-exposure model-electron throughput, the pixel stack, CFA response, pixel
integration, RGGB sampling and alias folding.  The random vector $Z$ is the
desired linear-sRGB image, not the output of a particular reconstruction
algorithm.  The prior, target, exposure and noise convention are fixed across
designs, and only the feasible pillar-width map $W$ changes.

The sole design objective is the camera-delivered target information
\begin{equation}
\boxed{
\mathcal I_{\mathrm{tar}}(W)
=\sum_{\theta}w_\theta I(Y_{W,\theta};Z_\theta)
=\sum_{\theta}w_{\theta}\,
\frac14\Big\langle\frac12\log_2
\frac{\det\mathbf{C}_Z(\nu)}
     {\det\mathbf{C}_{Z|Y,W}(\nu,\theta)}
\Big\rangle_{\nu}}
\label{eq:itar}
\end{equation}
in bit per raw pixel, with
\begin{equation}
\mathbf{C}_Z=T_a\mathbf{C}_{\mathrm{al}}T_a^\dagger,
\qquad
\mathbf{C}_{Z|Y,W}=T_aP_WT_a^\dagger,
\qquad T_a=T_a^{\rm prod}\ \text{from Eq.~\eqref{eq:target-prod}}.
\label{eq:targetposterior}
\end{equation}
The factor $1/2$ is the real-field convention and the factor $1/4$ converts a
four-pixel Bayer cell to a raw-pixel unit.  Field weights sum to unity.  The
$T_a$ has 12 rows and 64$K$ columns because the target is four 1\,\um{}
full-RGB phases, whereas the latent input contains sixty-four 0.25\,\um{}
aliases.

\subsection*{4.2 Posterior-entropy interpretation}

For the Gaussian model, Eq.~\eqref{eq:itar} is exactly
\begin{equation}
\mathcal I_{\mathrm{tar}}=H(Z)-H(Z\mid Y_W).
\label{eq:si-entropy-reduction}
\end{equation}
Thus the objective measures how much observing the raw model-electron frame
reduces the posterior volume of the desired color image.  It does not reward
latent spectral directions removed by $T$, and it is invariant to an invertible
change of coordinates within the same target space.  Changing only from XYZ to
linear sRGB therefore cannot change the objective. Changing the target task
requires changing its row space or rank.

\subsection*{4.3 Decoder independence inside the stated model}

Under joint Gaussianity, the conditional mean $\widehat Z=\mathbb E[Z\mid Y_W]$
is a sufficient statistic for $Z$.  Consequently
\begin{equation}
I(\widehat Z;Z)=I(Y_W;Z),
\label{eq:sufficient}
\end{equation}
and Eq.~\eqref{eq:itar} is a property of the camera channel, scene prior and
target, rather than information ``surviving'' a fixed demosaicker.  The
matched linear estimator used for posterior evaluation changes with $W$ because
its coefficients are rebuilt from $A_W$, but the prior, target and estimator class
remain fixed.  A learned nonlinear decoder under non-Gaussian natural-image
statistics could favor a different optical allocation.

Image reconstruction is used only for held-out, post-optimization checks.

\subsection*{4.4 I--MMSE and target-mode interpretation}

Let $\rho_i\in(0,1]$ be the generalized eigenvalues of the target posterior
against the target prior,
$\mathbf{C}_{Z|Y,W}v_i=\rho_i\mathbf{C}_Zv_i$, and define the corresponding
target-relevant mode signal-to-noise ratio $\gamma_i=\rho_i^{-1}-1$.  Before the
raw-pixel normalization, the information density is
\begin{equation}
\mathcal J_{\mathrm{tar}}(\nu,\theta)
=\frac12\sum_i\log_2(1+\gamma_i).
\label{eq:targetmodes}
\end{equation}
The scalar I--MMSE identity gives
$dI/d\mathrm{snr}=\tfrac12\mathrm{mmse}$ in nats.  Equivalently, the marginal
return of one target mode is
\begin{equation}
\frac{\partial\mathcal J_{\mathrm{tar}}}{\partial\gamma_i}
=\frac{1}{2\ln 2}\frac{1}{1+\gamma_i}.
\label{eq:marginal-information}
\end{equation}
Additional transfer therefore has less information value in a mode that is
already strong than in a weak target-relevant mode.  This is a balancing
tendency, not a claim that physical optimization holds total modal power fixed
or must equalize every mode.

\subsection*{4.5 Low-memory evaluation}

The sixty-four-alias state has dimension $64K=576$ at every coarse frequency.
A dense $576\times576$ aliased covariance over the full $104\times104$ grid
would consume tens of gigabytes before differentiation and is never
materialized. Write $A_q\in\mathbb C^{4\times K}$ and
$T_q\in\mathbb C^{12\times K}$ for the alias blocks of
Eqs.~\eqref{eq:operator-prod} and \eqref{eq:target-prod}, and retain the
independent $K\times K$ prior blocks $\mathbf{C}_q$. The required contractions are
\begin{align}
\mathbf{C}_Y &= \mathbf{C}_{\mathrm n}+\sum_q A_q\mathbf{C}_qA_q^\dagger,\\
\mathbf{C}_Z &= \sum_q T_q\mathbf{C}_qT_q^\dagger,\\
\mathbf{C}_{YZ} &= \sum_q A_q\mathbf{C}_qT_q^\dagger,\\
\mathbf{C}_{Y|Z} &= \mathbf{C}_Y-\mathbf{C}_{YZ}\mathbf{C}_Z^{-1}\mathbf{C}_{YZ}^\dagger.
\label{eq:measurement-space-cov}
\end{align}
The same local mutual information is then evaluated in measurement space,
\begin{equation}
I(Y;Z)=\frac12\log_2\frac{\det\mathbf{C}_Y}{\det\mathbf{C}_{Y|Z}}.
\label{eq:measurement-space-mi}
\end{equation}
All solves and factorizations are at most $12\times12$ or $4\times4$, and the
form is exactly equivalent to the target-posterior determinant under the stated
Gaussian model. Every accepted optimization checkpoint is selected by
$\mathcal I_{\mathrm{tar}}$ alone. Physical PSF, MTF, focal and charge
quantities are interpretation diagnostics, while image metrics are
post-optimization tests.

% =====================================================================
\snote{5: Target-mode allocation and the restricted transfer-energy limit}
\label{sn:modes}

\subsection*{5.1 The whitened operator and its singular values}

Treating the metalens and the Bayer mosaic as one matrix at each spatial
frequency shows what the width map does to the sensor. Define the diagnostic
whitened operator
\begin{equation}
\widetilde A_W(\nu)=\mathbf{C}_{\mathrm n}^{-1/2}A_W(\nu)\,\mathbf{C}_{\mathrm{al}}^{1/2},
\label{eq:whiten}
\end{equation}
which removes the units of the noise and of the scene statistics. Its singular
values $\sigma_1(\nu)\ge\sigma_2(\nu)\ge\cdots$ are the signal-to-noise
amplitudes of independent measurement modes. They are useful diagnostics, but
the design objective is target-conditioned. The generalized
eigenvalues $\rho_i\in[0,1]$ of the target posterior against the target prior,
$\mathbf{C}_{z|y}v_i=\rho_i\mathbf{C}_z v_i$, are the fraction of posterior variance
remaining in each color--spatial target mode, and $1-\rho_i$ is the
\emph{target-mode resolution}, the fraction of that mode's variance the camera
recovers. Equivalently, $\gamma_i=\rho_i^{-1}-1$ is its target-relevant modal
signal-to-noise ratio. Per coarse frequency, before the raw-pixel and field
averages in Eq.~\eqref{eq:itar},
\begin{equation}
\mathcal J_{\mathrm{tar}}(\nu)
=-\frac12\sum_i\log_2\rho_i
=\frac12\sum_i\log_2(1+\gamma_i).
\label{eq:modes}
\end{equation}

\subsection*{5.2 The four-measurement rank ceiling}

The measurement operator $A_W^{\rm prod}(\nu)$ is $4\times64K$ and
the 1\,\um{} full-RGB target operator $T_a^{\rm prod}(\nu)$ is
$12\times64K$. The camera rank is therefore at most four, so at each coarse
frequency at most four of the twelve target directions can receive independent
measurement information, whatever the optic. The four raw phases set that
ceiling, and the design freedom changes which target-relevant combinations receive
signal-to-noise.

\subsection*{5.3 Squared transfer energy as a restricted limit}

For one scalar, alias-free channel
\begin{equation}
Y(\nu)=H_W(\nu)X(\nu)+N(\nu),
\end{equation}
the Gaussian information is
\begin{equation}
I(X;Y)=\frac12\sum_{\nu}\log_2\!\left[
1+\frac{S_X(\nu)}{S_N(\nu)}|H_W(\nu)|^2\right].
\label{eq:si-scalar-information}
\end{equation}
If the scene and noise spectra are white, the channel is scalar, aliases and
target projection are absent, and the operating point is sufficiently low SNR
that $\log(1+x)\simeq x$, then
\begin{equation}
I(X;Y)\ \propto\ \sum_{\nu}|H_W(\nu)|^2.
\label{eq:mtf-limit}
\end{equation}
A squared transfer-energy objective is therefore recovered as a physically
meaningful restricted approximation. Conventional MTF volume usually integrates
the first power $|H|$ after DC normalization. It is related to, but is not
algebraically identical to, the squared absolute-transfer limit in
Eq.~\eqref{eq:mtf-limit}. The full camera objective additionally retains spectral
coupling, colored scene statistics, signal-dependent noise, non-DC-normalized
exposure-calibrated throughput,
the target projection and sampling aliases.

The same connection appears in matrix form.  After replacing the usual
DC-normalized modulus by absolute squared transfer, inserting scene weighting and
whitening the noise,
$\operatorname{tr}(\widetilde A_W\widetilde A_W^\dagger)=\sum_i\sigma_i^2$.
The low-SNR expansion of the log determinant is proportional to this modal total.
Away from that restricted limit, total modal power does not specify its allocation
or orientation relative to the target.

\medskip\noindent\textbf{(i) Allocation at fixed modal total.} Fix the total
$T=\sum_{i=1}^{n}\sigma_i^2$. By concavity of $x\mapsto\log(1+x)$ and Jensen's
inequality~\citep{cover2006},
\begin{equation}
\sum_{i=1}^{n}\log_2\!\big(1+\sigma_i^2\big)\ \le\ n\log_2\!\Big(1+\frac{T}{n}\Big),
\label{eq:jensen}
\end{equation}
with equality if and only if all $\sigma_i^2$ are equal.  At fixed modal total,
the information is therefore larger for a balanced allocation.  The physical
optimization does not fix that total, so Eq.~\eqref{eq:jensen} explains a tendency
rather than imposing a universal equalization rule.

\medskip\noindent\textbf{(ii) Effect of channel collinearity.} A per-wavelength scalar
MTF depends only on the magnitudes of the individual channel transfers, that is on
the \emph{diagonal} of $M\mathbf{C} M^{\dagger}$. The information depends on the whole
matrix. Two operators with identical diagonals can have Gram matrices of rank one
(all three channel transfers collinear, so the camera has no color discrimination
whatever) or of full rank (orthogonal transfers, full discrimination), and they
carry identical MTF volume. This can be quantified. Suppose
a single scene mode $s$ of variance $\sigma_s^2$ leaks into all three
channels through overlapping passbands, $H_a(\nu)=\gamma_a h(\nu)$, so
$m_a=\gamma_a h s+n_a$. The true joint channel is rank one,
\begin{equation}
I_{\mathrm{joint}}=\tfrac12\log_2\!\Big(1+\sigma_s^2|h|^2\sum_a \gamma_a^2/\sigma_a^2\Big),
\label{eq:rank1}
\end{equation}
whereas a diagonal per-channel sum reports three separate logarithms. With equal
weights $\gamma_a=\gamma$, $\sigma_a=\sigma$ and $x=\sigma_s^2\gamma^2|h|^2/\sigma^2$,
\begin{equation}
\frac{I_{\mathrm{diag}}}{I_{\mathrm{joint}}}
=\frac{3\log_2(1+x)}{\log_2(1+3x)}\ \xrightarrow[\ x\gg1\ ]{}\ 3 ,
\label{eq:rankinfl}
\end{equation}
so at high signal-to-noise a diagonal accounting exceeds the joint value by one
factor per collinear copy, because it does not diagonalize the shared mode.

\medskip\noindent Two further effects follow from the same source. A
\emph{DC-normalized} transfer has unit DC
by construction, so two designs with different calibrated model-electron throughput (hence
different shot-limited signal-to-noise at a common exposure) receive identical
credit, whereas the exact operator keeps calibrated gain in $M$. A transfer
functional integrated over the full pixel-frequency grid also treats every
frequency as independently observed, folding no aliases, whereas the operator of
Eq.~\eqref{eq:operator-prod} and the block prior of Eq.~\eqref{eq:sigmaal}
perform the fold explicitly. Evaluated on the matrix-valued operator, the
target information accounts jointly for total, allocation, collinearity,
calibrated gain and alias folding.  A large scalar MTF volume is therefore not, by
itself, sufficient to determine camera-delivered target information.  The
matched-budget MTF comparison is reported in Supplementary Note~7 and in the
main text.

% =====================================================================
\snote{6: Optimization protocol: parameterization, schedule, field subsampling, checkpoint selection, and the measurement convention}
\label{sn:opt}

\subsection*{6.1 What varies and what does not}

Material, pillar height, lattice pitch, aperture diameter, numerical aperture,
focal length, object distance, the square-pillar library and its width bounds, the
CFA and source spectra, the exposure calibration, the noise model, the scene prior
and the field quadrature are all fixed (Supplementary Table~\ref{tab:constants}).
The only design variable is the width map $W(x,y)$.

\subsection*{6.2 Differentiability and parameterization}

The path is differentiable through width interpolation of the
full-Jones RCWA response, vectorial propagation, sliding photosite aperture, the
64-alias operator of Eq.~\eqref{eq:operator-prod} and the alias-block
contractions of \S4.5. The PyTorch implementation~\citep{paszke2019} is verified
by a finite-difference gradient check over the Q64 bridge.

The optimization varies widths on one mirror quadrant of the
$720\times720$ lattice, expanded to the full lattice by the mirror
symmetry and box-projected onto the library-supported interval. No
spatial filter is applied. The exact active-post count is taken from the stored
circular aperture mask, and the number of independent optimization variables is
the active mirror-quadrant site count, 101{,}012. The
$720\times720=518{,}400$ number is only the square-grid site count. Inactive
sites are neither posts nor optimization variables.

\subsection*{6.3 Single-objective constrained schedule}

The optimization is
\begin{equation}
W^\star=\arg\max_{W\in\mathcal F}\mathcal I_{\mathrm{tar}}(W),
\label{eq:constrained-itar}
\end{equation}
where $\mathcal F$ enforces the width interval and the
mirror-quadrant parameterization. Width bounds are applied by exact box
projection. No spatial filter, smoothness penalty or fabrication barrier is
added to Eq.~\eqref{eq:constrained-itar}, and the scalar loss passed to the optimizer
is only $-\mathcal I_{\mathrm{tar}}$.

The symbol $\mathcal I_{\mathrm{tar}}$ in Eq.~\eqref{eq:constrained-itar}
denotes one implementation path only:
the response is assembled into $A_W^{\rm prod}$ by
Eq.~\eqref{eq:operator-prod}, its sixty-four alias blocks are streamed through
the contractions of Eq.~\eqref{eq:measurement-space-cov}, and the scalar is
evaluated from the measurement-space determinant in
Eq.~\eqref{eq:measurement-space-mi}.

The run uses one stage of projected Adam~\citep{kingma2015adam} for 300
mirror-quadrant steps under a cosine schedule from $1.6\times10^{-2}$ to
$10^{-5}$, with four stochastic field samples per step and deterministic
all-25-field validation every ten steps. The same fixed hyperbolic reference
seed initializes every arm, so the native objective is the only intentional
difference between them. Stochastic field sampling uses seed 20260805.

\subsection*{6.4 Initialization and local-optimum scope}

Gradient optimization is local, so the starting point matters. Optimization
starts from the stored \SI{600}{nm} hyperbolic reference projected onto
the feasible set by box projection and mirror-quadrant symmetrization, and the
same seed initializes every arm. Any resulting comparison is confined to that
basin: the layout of Fresnel-zone boundaries and per-wavelength branch
assignments contains effectively discrete structure that a local gradient need
not cross. No global-optimum claim is made.

The comparison reference is the corrected \SI{600}{nm} hyperbolic map because,
among the nine design wavelengths, \SI{600}{nm} places the R-channel focal peak
closest to the sensor plane at $F=346.7\,\mu$m, giving the smallest
on-axis red-channel spot and the highest single-wavelength-focused conventional
baseline for this geometry. A \SI{600}{nm} prescription also carries near-maximal
target information before any optimization, sitting on the top plateau of the
initial nine-wavelength values within
\num{0.001}~bit per raw pixel of the
maximum, so it is the best-justified single starting camera in the objective's
own terms and not a tuned selection.

\subsection*{6.5 Field quadrature}

Every target-information value and the design objective itself is a
weighted mean over one field quadrature. The quadrature is therefore specified in
full below.

The field of view is a \emph{square}. The filtered width geometry passes the
explicit D4 axis-reflection\slash transpose gate, but no reduction of the complete
polarized-Jones, fixed-RGGB and target-prior operator is assumed from that
geometric fact alone. The quadrature therefore covers a full first
quadrant rather than inferring the 45--90$^\circ$ fields from the 0--45$^\circ$
ones. It is built as a product of a radial rule and an azimuthal rule.

\emph{Physical square-field boundary.} At the reference wavelength
$\lambda_{\rm ref}=540$\,nm, the largest allowed sensor-plane chief-ray radius
at azimuth $\phi$ is
\begin{equation}
s_{\max}(\phi)=\min\!\left[
\frac{N_x\Delta_s/2-\delta}{|\cos\phi|},
\frac{N_y\Delta_s/2-\delta}{|\sin\phi|}
\right],
\label{eq:rmax}
\end{equation}
where a zero denominator contributes no bound, $\Delta_s=0.25$\,\um{} and
$\delta=5$\,\um{} is a retained-PSF margin. Conservation of tangential wave
vector then gives
\begin{align}
\theta_{\rm out,max}(\phi)&=\tan^{-1}\!\frac{s_{\max}(\phi)}{F},\\
\theta_{\rm in,max}(\phi)&=\sin^{-1}\!\left[
\frac{n_{\rm out}(\lambda_{\rm ref})}{n_{\rm in}(\lambda_{\rm ref})}
\sin\theta_{\rm out,max}(\phi)\right],\\
r_{\max}(\phi)&=\min\!\left[
z_{\rm obj}\tan\theta_{\rm in,max}(\phi),
\frac{L_x/2}{|\cos\phi|},\frac{L_y/2}{|\sin\phi|}
\right].
\label{eq:squarescale}
\end{align}
Thus the square sensor, the square object support and the substrate-to-air
refraction all constrain the field before optimization.

\emph{Radial and azimuthal rules.} Five azimuths, $0$, $22.5$, $45$, $67.5$
and $90^\circ$, carry composite Simpson weights
$w_\phi=\{1,4,2,4,1\}/12$. With $n=5$ rings, the node radius is the equal-area
midpoint
\begin{equation}
r(i,\phi)=r_{\max}(\phi)\sqrt{\frac{i+\tfrac12}{n}},
\qquad i=0,\dots,n-1 .
\label{eq:ringradius}
\end{equation}
The unnormalized node weight is
$w_\phi r_{\max}(\phi)^2/n$ and the 25 weights are normalized to unit sum.
The field angle recorded in the operator is the substrate-side
$\theta(i,\phi)=\tan^{-1}[r(i,\phi)/z_{\rm obj}]$.

The same 540-nm common chief-ray translation is applied to every wavelength,
so wavelength-dependent lateral color is not recentered away. Every node passes
a pupil-carrier Nyquist check, an air-side first-lattice-order evanescence check for the
zeroth-order-only forward and a direct PSF centroid/border-energy test.

\emph{The quadrature contains no on-axis node.} The equal-area midpoint rule of
Eq.~\eqref{eq:ringradius} places the first ring at
$r(0,\phi)=r_{\max}(\phi)/\sqrt{10}$ rather than at zero, so the innermost value
is not an on-axis value. The near-degenerate
pattern of the 0$^\circ$/90$^\circ$ and 22.5$^\circ$/67.5$^\circ$ pairs within
a ring is the signature of a finite field angle under square symmetry rather
than of numerical noise about zero.

\subsection*{6.5.1 First-quadrant sufficiency under mirror-quadrant symmetry}

The objective is evaluated over 25 first-quadrant fields, yet the camera
covers the full field of view. The first-quadrant restriction is
self-consistent for the following reasons. (i) The width map has D2 mirror
symmetry by construction (mirror-quadrant parameterization, \S6.1). (ii) The
field-local sensor shift translates every field's point-spread function to a
common local CFA patch, so the PSF-to-mosaic registration is identical at
mirror-related fields. (iii) Under the stationary periodic prior, a
global CFA phase relabel is an invertible reindexing of $X$ and $Z$ that
leaves $I(Z;Y_W)$ invariant. The three conditions together imply that
mirror-related fields contribute equal per-field target information, and
averaging over one quadrant reproduces the full-field mean. Spectral-PSF mirror
parity is checked numerically at each checkpoint. Under a non-stationary prior or
a position-tied metric the first-quadrant restriction would be an approximation
rather than an identity.

\subsection*{6.6 Field subsampling and checkpoint selection}

The schedule draws four fields per step with replacement by their
normalized quadrature weights and averages their per-field objective values, an
unbiased stochastic estimate of the full field average. Each run uses 300 steps,
with checkpoints selected by deterministic evaluation over all 25 fields every
ten steps and an identical five-step full-field refinement at rate $10^{-4}$ for
each arm.

\subsection*{6.7 The measurement convention}

Every design is evaluated from its stored $720\times720$ full map through one
common evaluator, with no refitting of a reduced profile and no annular
averaging. The mirror-quadrant parameterization belongs to optimization, and the
stored expanded map is the physical state that is rescored. The objective is the
quadrature-weighted field mean over the 25 fields.

% =====================================================================
\snote{7: MTF-volume baseline construction}
\label{sn:baselines}

The in-house broadband MTF-volume control is motivated by the published
broadband MTF-volume objective~\citep{froch2025} and is used as the
representative restricted proxy derived in \S5.3. It is not an exact
reproduction of that crystalline-silicon optic, its wavelength and field grid,
or its Wiener/BM3D reconstruction pipeline. A fair comparison requires the same
meta-atom library and common forward
model, wavelength and field sampling, degrees of freedom, feasible-set
parameterization, optimization step budget and checkpoint protocol as the
information design.  The retained map must also demonstrably improve its own
native MTF objective before all maps are rescored by the common camera
evaluator.

The matched run uses the feasible set, seed, field and wavelength
quadrature and step budget, and raises its native DC-normalized MTF volume from
the shared hyperbolic seed's \num{0.0491} to \num{0.0711} over the recorded
optimization stage, an improvement of \num{44.6}\%. One common learning rate,
$1.6\times10^{-2}$, is for every arm, so the arms differ in native
objective alone and no per-arm rate selection enters the comparison.

Only a matched run separates the effect of the objective from geometry or
optimization budget: Eq.~\eqref{eq:mtf-limit} states when the proxy approximates
information, while the experiment measures how far this coupled color camera
lies from that limit.

% =====================================================================
\snote{8: Imaging pipeline, reconstruction, and metric definitions}
\label{sn:imaging}

\begin{figure}[t]
\centering
\includegraphics[width=\linewidth]{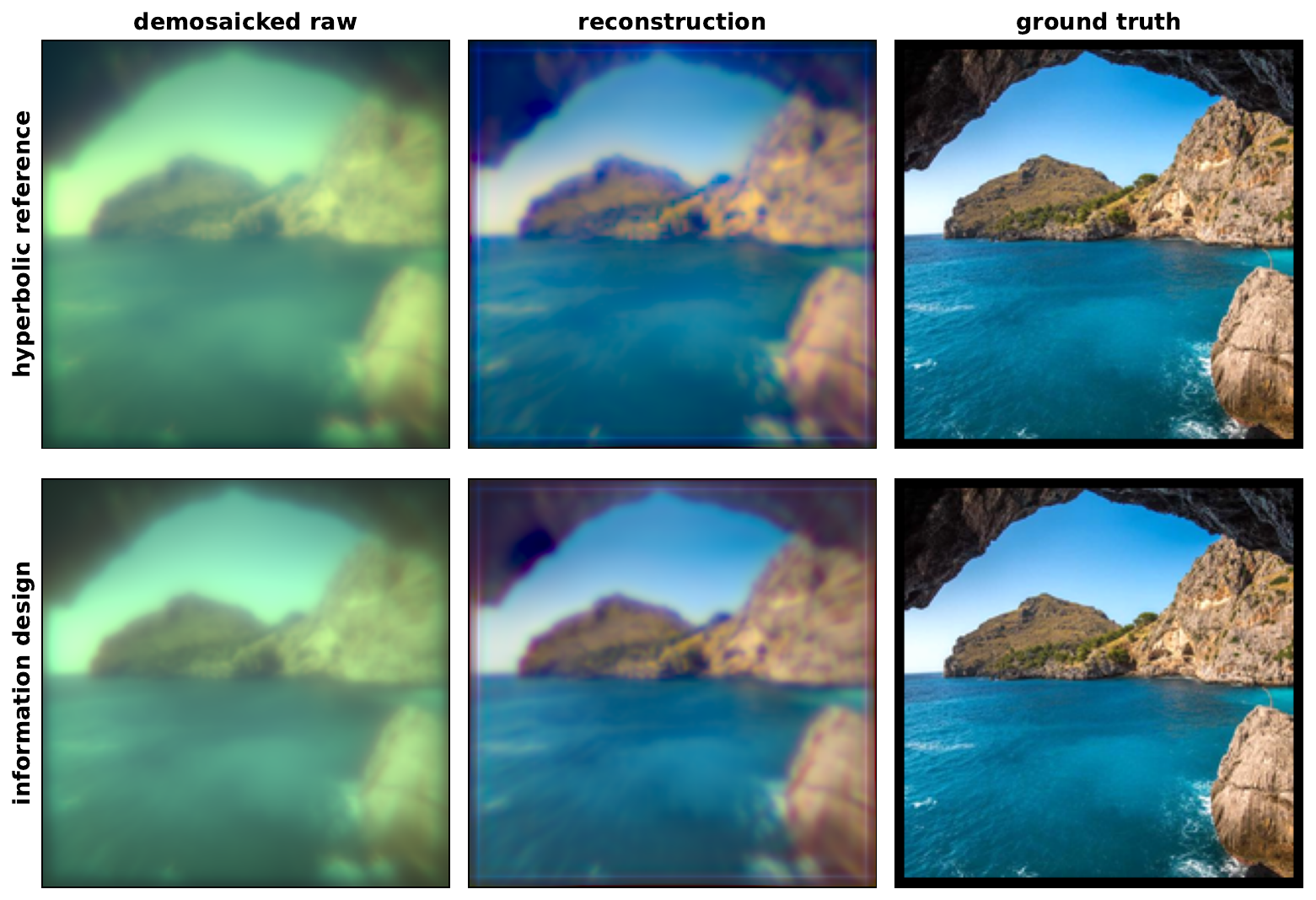}
\caption{\textbf{Imaging pipeline on the landscape scene (SiN).} Left, the
bilinearly demosaicked raw sensor frame before reconstruction (own-percentile
display, color-uncalibrated). Middle, the field-dependent Wiener reconstruction
after the fixed analytic color calibration. Right, the ground truth. The
demosaicked raw is a soft, aliased, green-dominated capture. The reconstruction
recovers structure and color, and the information design (bottom) is sharper and
more faithful than the hyperbolic reference (top).}
\label{sfig:rawdemosaic}
\end{figure}

The target-information objective of Supplementary Note~4 is computed on a local,
periodic operator. The imaging evaluation is deliberately \emph{not}: it renders
whole scenes through the field-varying PSF of each design across the full square
field of view, adds the Poisson and read-noise surrogates on the
model-electron scale, and reconstructs
with one fixed practical protocol that is identical for every design. This note
specifies that pipeline exactly.

\subsection*{8.1 Direct full-scene forward propagation}

The held-out scene is rendered by direct full-scene propagation. Every object
point of the scene is mapped through the full-Jones forward model of the design,
the same model and calibrated transfer that define the target-information
objective, and the resulting model-electron fields are accumulated on the RGGB
photosite grid. No field is approximated by a mirrored or rotated field.
Geometric symmetries are used only for computational acceleration. All color
channels share one geometric chief-ray origin at each field, so lateral color
is preserved.

A field-dependent RGB point-spread-function bank is sampled separately on a
signed $8\times8$ Cartesian field grid from the same forward model and calibrated
transfer. It enters only the Wiener reconstruction decoder of \S8.3 and does not
form the raw image. The decoder kernel size is selected after convergence of the
reconstruction metrics at 101, 201 and 401 pixels.

\subsection*{8.2 Exposure, noise and color calibration}

The primary imaging comparison uses one fixed exposure for every design, with
the model-electron calibration, Poisson draw convention and 1.5\,\eminus{} rms
read noise held fixed.  This preserves the throughput and shot-noise differences
that enter $\mathcal I_{\mathrm{tar}}$.  Equal-brightness renders, in which each
design is independently exposed to a common center-field mean, are a secondary
blur and color-allocation diagnostic.

Color calibration is a per-design analytic white-balance and color matrix
constructed from a predeclared reference-camera DC mapping before
rendering. For each design it uses that design's own on-axis
spectral DC response, the covariance-weighted right inverse of the
linear-sRGB target and analytic constant-channel probes through the fixed
bilinear and Wiener decoder. The construction has no raster-scene input, so each
$3\times3$ matrix is analytic rather than fitted. It is fixed before the design
sees any test scene and applied unchanged across noise realizations. No per-scene
fitting to the test ground truth is used.

\subsection*{8.3 Reconstruction used for the imaging check}

The RGGB raw frame is demosaicked by a fixed bilinear rule, and each color
channel is reconstructed with the same field-dependent Wiener protocol.
Reconstruction coefficients are not part of Eq.~\eqref{eq:itar}. Metrics are
computed from the saved arrays. The Gaussian conditional mean in Supplementary
Note~4 is used only to evaluate the closed-form target posterior and is not the
image-rendering algorithm.

\subsection*{8.4 RGGB measurement path}

The forward render applies wavelength-dependent optical blur and pixel
integration before the binary RGGB sampling mask, then adds
model-electron-domain noise only at the measured raw samples. Demosaicking
occurs only after this raw measurement has been saved. The same scene, exposure
and paired-noise protocol is used for all designs. The RGGB lattice is one
concrete measurement operator inside $A_W$, not the name or definition of the
design objective.

\subsection*{8.5 Scene}

The primary imaging evaluation uses one held-out landscape scene, which is
neither the optimization scene nor the ensemble prior.

\subsection*{8.6 Metric definitions}

Primary metrics are computed after the independently constructed, fixed analytic
color calibration of \S8.2. The same post-calibration domain is used for every
metric: both target and reconstruction are clipped once to $[0,1]$ in linear
RGB, the clipped fractions are recorded, and PSNR uses a fixed data range of
one. No metric-specific pre-clipping is permitted. They are
\begin{itemize}
\item \textbf{PSNR}, on calibrated linear RGB, for overall reconstruction fidelity.
\item \textbf{$\Delta E_{00}$}, CIEDE2000 after conversion through CIE XYZ to
CIELAB under the D65 white point, using $k_L=k_C=k_H=1$, for color
fidelity.
\item \textbf{S-CIELAB}, under the fixed viewing convention, for spatially weighted
color fidelity.
\end{itemize}
S-CIELAB uses 10 pixels per degree. Its three opponent filters use the following
Gaussian $(w,\sigma_{\rm deg})$ pairs:
\begin{equation*}
\begin{aligned}
F_1 &: (0.921,0.0283),\ (0.105,0.133),\ (-0.108,4.336),\\
F_2 &: (0.531,0.0392),\ (0.330,0.494),\\
F_3 &: (0.488,0.0536),\ (0.371,0.386).
\end{aligned}
\end{equation*}
Each filtered channel is divided by its weight sum. The viewing scale, kernels,
white point, RGB-to-XYZ matrix, clipping policy and data range are fixed.
Euclidean $\Delta E_{76}$ is a separately named diagnostic and is not the primary
color metric.

\subsection*{8.7 External inputs}

Several quantities used throughout are measured or estimated data rather than
model parameters, and a reader cannot reproduce the pipeline without them. Each
is provided with the code.

\begin{itemize}
\item \emph{Scene spectral covariance shape $C$} ($9\times9$, \S3.2). An
empirical dimensionless shape normalized to unit mean XYZ variance. The spatial
PSD of Eq.~\eqref{eq:psd} supplies the only contrast variance.
\item \emph{Color-filter, quantum-efficiency and source curves.} These are the
three spectral factors of Eq.~\eqref{eq:eweight}: the Samsung Galaxy S20
mobile color-filter curves measured by Tominaga et al.\citep{tominaga2021},
sampled at the nine optimization wavelengths. The quantum
efficiency is held at unity throughout, so every model-electron count is a
relative rather than an absolute count.
\item \emph{Meta-atom lookup table.} The full-Jones response table, with its
index models, grids and truncation order, is specified in Supplementary
Table~\ref{tab:library} and provided as arrays.
\item \emph{Reconstruction calibration.} Wiener parameters and the fixed color
calibration are constructed from the analytic reference-camera DC
mapping and provided with the code.
\item \emph{Evaluation scene.} The held-out landscape test image (\S8.5).
\end{itemize}

% =====================================================================
\snote{9: Extended results}
\label{sn:extended}

This note reports additional analyses of information decomposition, axial focal
structure, wavelength-resolved transfer and scene robustness.

\begin{figure}[t]
\centering
\includegraphics[width=\linewidth]{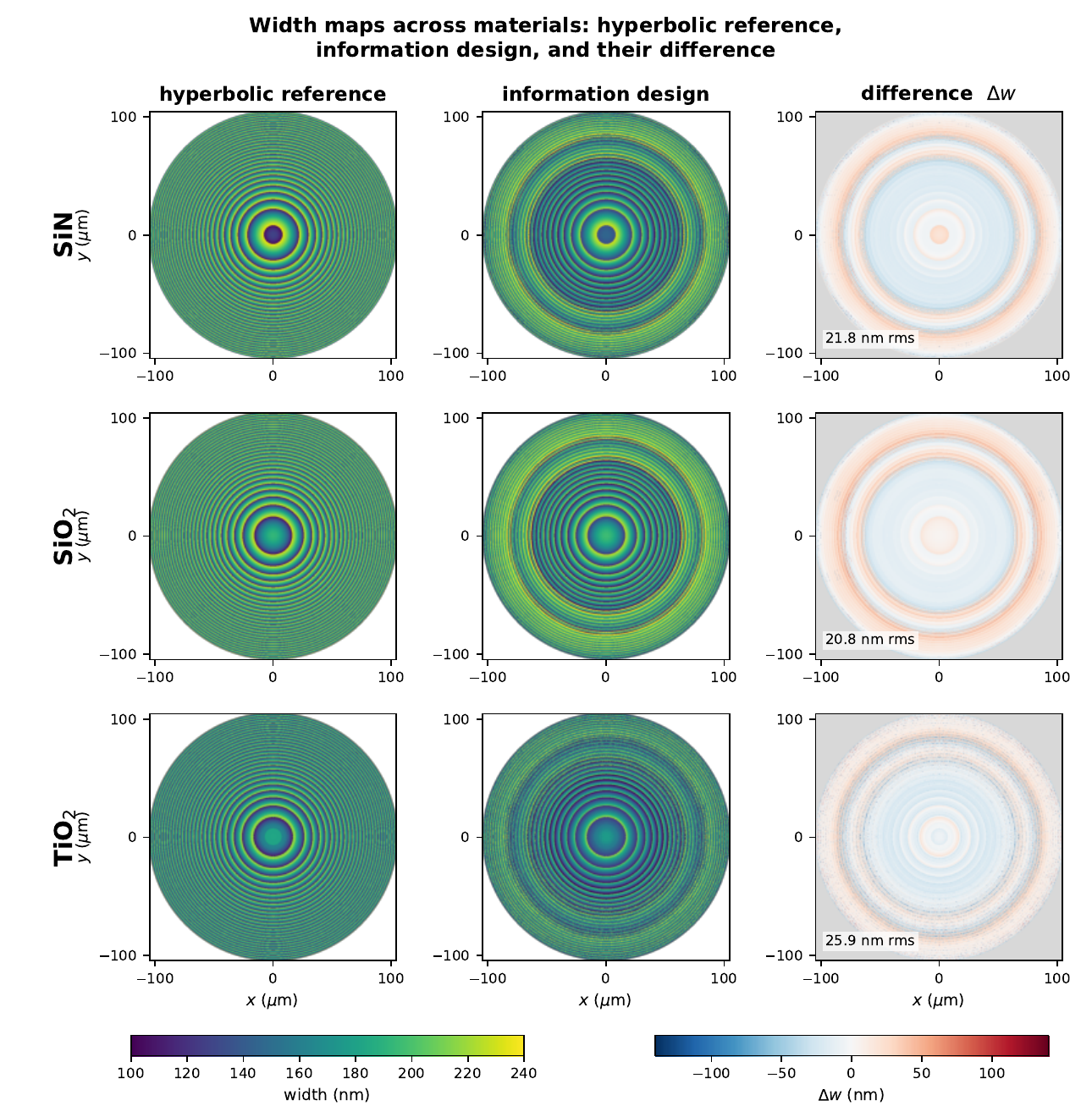}
\caption{\textbf{Width maps across meta-atom materials.} For SiN, SiO$_2$ and
TiO$_2$ (rows), the hyperbolic reference width map, the information-design width
map, and their difference (information minus reference). The difference is
\SI{21.8}{nm}, \SI{20.8}{nm} and \SI{25.9}{nm} rms over the active aperture for
SiN, SiO$_2$ and TiO$_2$ respectively. Each material's information design
redistributes the width assignment by a comparable amount while keeping the same
physical library and pitch.}
\label{sfig:widthmaps}
\end{figure}

\subsection*{9.1 Total and target information decomposed}

The chain rule $I(X;Y)=I(Z;Y)+I(X;Y\mid Z)$ separates the total captured
information $\mathcal I_{\mathrm{tot}}=I(X;Y_W)$ into the delivered target
information $\mathcal I_{\mathrm{tar}}=I(Z;Y_W)$ and a target-null conditional
remainder $I(X;Y_W\mid Z)$, the information the raw measurement carries about
metameric spectral variation that leaves the color target unchanged.
Supplementary Table~\ref{tab:decomp} gives this decomposition for the hyperbolic
reference, the MTF-volume control and the information design. Resolved by field,
the MTF-volume control falls below the reference in all \num{25} fields on both
information terms, and on the per-field mean it suppresses the target-null
remainder by \num{63.3}\% against \num{17.5}\% for the delivered target
information.

\begin{table}[htbp]
\centering
\caption{\textbf{Total, target and conditional information at the fixed common
exposure.} 25-field quadrature-weighted values from the decomposition,
with the mean raw-pixel charge $(R+2G+B)/4$ on the same quadrature. Every design
is scored by one evaluator from its stored width map. Columns are rounded
independently. On the unrounded values the chain rule holds to \num{1.1e-7} bit
on every row.}
\label{tab:decomp}
\begin{tabular}{lcccc}
\hline
Design & $\mathcal I_{\mathrm{tot}}$ & $\mathcal I_{\mathrm{tar}}$ & $I(X;Y_W\mid Z)$ & charge (model-e\textsuperscript{-}/raw px)\\
\hline
Hyperbolic reference & 0.5650 & 0.5059 & 0.0591 & 724.0\\
MTF-volume control & 0.4384 & 0.4180 & 0.0204 & 274.8\\
Information design & 0.6304 & 0.5860 & 0.0444 & 624.4\\
\hline
\end{tabular}
\end{table}

\subsection*{9.2 Axial focal structure}

\begin{figure}[htbp]
\centering
\includegraphics[width=\linewidth]{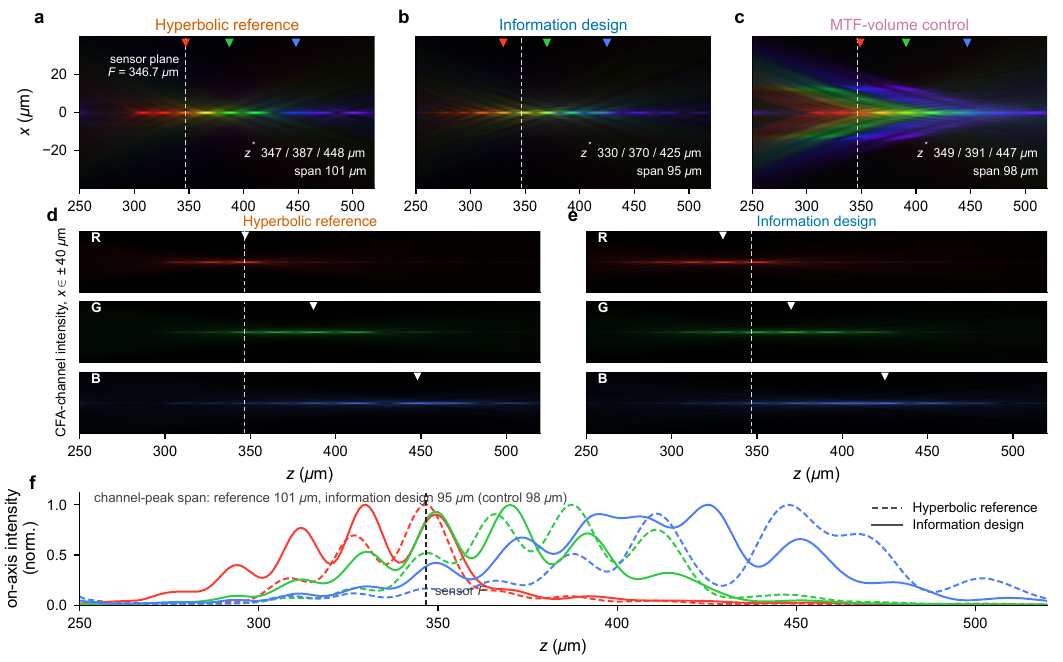}
\caption{\textbf{Longitudinal chromatic structure and the channel foci.}
\textbf{a}--\textbf{c} True-color longitudinal composites for the hyperbolic
reference, information design and MTF-volume control within the saved
\SIrange{250}{520}{\micro\meter} window, with the sensor plane at
$F=\SI{346.7}{\micro\meter}$ marked. The color-filter channel intensity peaks
sit at $(347,\,387,\,448)$, $(330,\,370,\,425)$ and
$(349,\,391,\,447)$~\si{\micro\meter} for the reference, information and
MTF-volume designs, with channel-peak spans of \SI{101}{\micro\meter},
\SI{95}{\micro\meter} and \SI{98}{\micro\meter}. \textbf{d},\textbf{e} R, G and B
channel longitudinal sections for the reference and the information design.
\textbf{f} On-axis per-channel profiles, reference dashed against information
design solid, each peak-normalized for display. All three designs spread their
channel-intensity global maxima across about \SI{100}{\micro\meter}, but each
channel keeps a local intensity maximum within a few micrometers of the sensor
plane, and the information design balances these near-sensor values across
channels more evenly than the reference or the control.}
\label{fig:xz}
\end{figure}

Longitudinal focus is inspected after optimization for the same three maps used
in the main comparison: the hyperbolic reference, the matched-budget MTF-volume
design and the information design. For each wavelength, the axial peak is read
directly from the saved unrecentered $x$--$z$ field diagnostic
(Supplementary Fig.~\ref{fig:xz}). Detector-plane point-spread functions are
read from the same caches. This comparison is descriptive. No
dedicated focal-spread-control map was produced, so the data do not support a
causal claim that the information optimum differs from a separately optimized
focal-coincidence solution.

At the saved $z$ sample nearest the sensor plane (\SI{347}{\micro\meter} against
$F=\SI{346.7}{\micro\meter}$), the on-axis, quadrature-masked ($|x|\le\SI{2}{\micro\meter}$),
CFA-weighted intensity of each channel is at or below that channel's own global
axial maximum for every design. As a fraction of that maximum, the reference gives
\num{1.00}, \num{0.52} and \num{0.17} for R, G and B, the MTF-volume control
gives \num{0.98}, \num{0.55} and \num{0.26}, and the information design gives
\num{0.85}, \num{0.87} and \num{0.40}. Each design keeps a genuine local
intensity maximum within \SI{4}{\micro\meter} of the sensor plane in every
channel, and the information design's near-sensor values are the most balanced
of the three, with the highest green and blue fractions, so the spread of the global maxima in
Supplementary Fig.~\ref{fig:xz} does not imply a correspondingly spread
in-focus response at the sensor itself.

\subsection*{9.3 Wavelength-resolved transfer}

\begin{figure}[htbp]
\centering
\includegraphics[width=\linewidth]{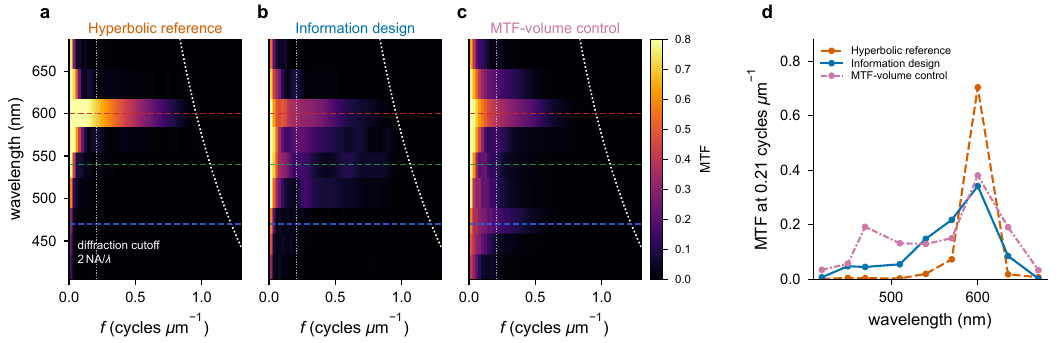}
\caption{\textbf{Wavelength-resolved transfer.} \textbf{a}--\textbf{c}
Radially averaged magnitude of the raw OTF normalized at DC, as a function of
wavelength and spatial frequency for the hyperbolic reference, information
design and MTF-volume control on a shared scale, with the color-filter
centroids dashed and the diffraction cutoff $2\,\mathrm{NA}/\lambda$ at the
exact geometric air NA \num{0.2873} dotted.
\textbf{d} The cut at \SI{0.21}{\cycle\per\micro\meter} for all three designs
at the nine stored wavelengths, plotted at their true non-uniform positions
with no interpolation across wavelength. The reference concentrates transfer
on one \SI{600}{nm} ridge, reaching \num{0.704}. The information design
halves that ridge to \num{0.342} and the MTF-volume control sits between at
\num{0.381}, while the information design holds \num{0.148} and \num{0.218}
at \num{540} and \SI{570}{nm} where the reference holds \num{0.019} and
\num{0.073}. Transfer is redistributed toward the previously starved bands.}
\label{fig:mtf}
\end{figure}

The reference concentrates transfer on a single \SI{600}{nm} ridge. The
information design redistributes it toward the previously starved \num{540} and
\SI{570}{nm} bands (Supplementary Fig.~\ref{fig:mtf}). This is the transfer-domain
view of the spot-size reallocation reported in the main text.

\subsection*{9.4 Imaging comparison including the MTF-volume control}

\begin{figure}[htbp]
\centering
\includegraphics[width=\linewidth]{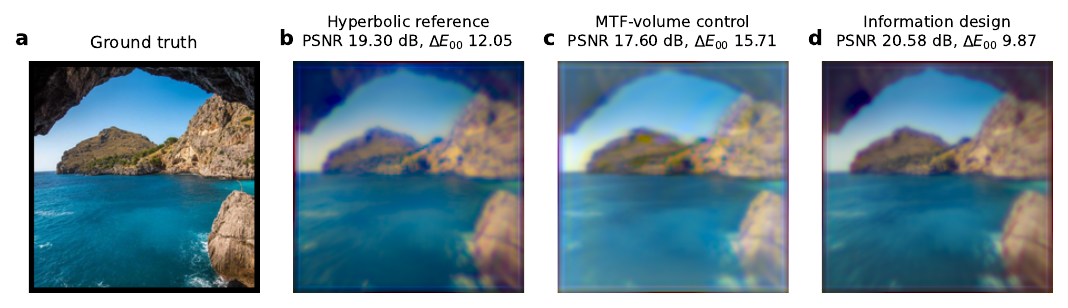}
\caption{\textbf{Held-out scene reconstruction including the MTF-volume
control.} \textbf{a} Ground truth. \textbf{b}--\textbf{d} Noiseless
reconstructions from the hyperbolic reference, the MTF-volume control and the
information design, each under its own independent analytic color
calibration, with PSNR and $\Delta E_{00}$ against the ground truth.}
\label{fig:mtfimaging}
\end{figure}

The main-text imaging figures compare the hyperbolic reference against the
information design because the MTF-volume control is
introduced in the Results as a test of a published transfer-domain criterion
rather than as a candidate design in its own right. For completeness,
Supplementary Fig.~\ref{fig:mtfimaging} reconstructs the same held-out
landscape scene from all three saved width maps under the identical pipeline
of \S8.3: direct full-scene propagation, demosaicing, the fixed
field-dependent Wiener step and each design's own independent analytic color
calibration, with no per-scene refitting. The MTF-volume control gives
\SI{17.60}{\decibel} PSNR and $\Delta E_{00}=15.71$, against
\SI{19.30}{\decibel} and \num{12.05} for the reference and
\SI{20.58}{\decibel} and \num{9.87} for the information design, matching the
values reported in the main text. The MTF-volume control is the weakest of
the three designs on both metrics, consistent with the delivered target
information and charge already reported for it in the main text and in
Supplementary Table~\ref{tab:decomp}.

\subsection*{9.5 Scene robustness across held-out natural images}

\begin{figure}[tb]
\centering
\includegraphics[width=0.88\linewidth]{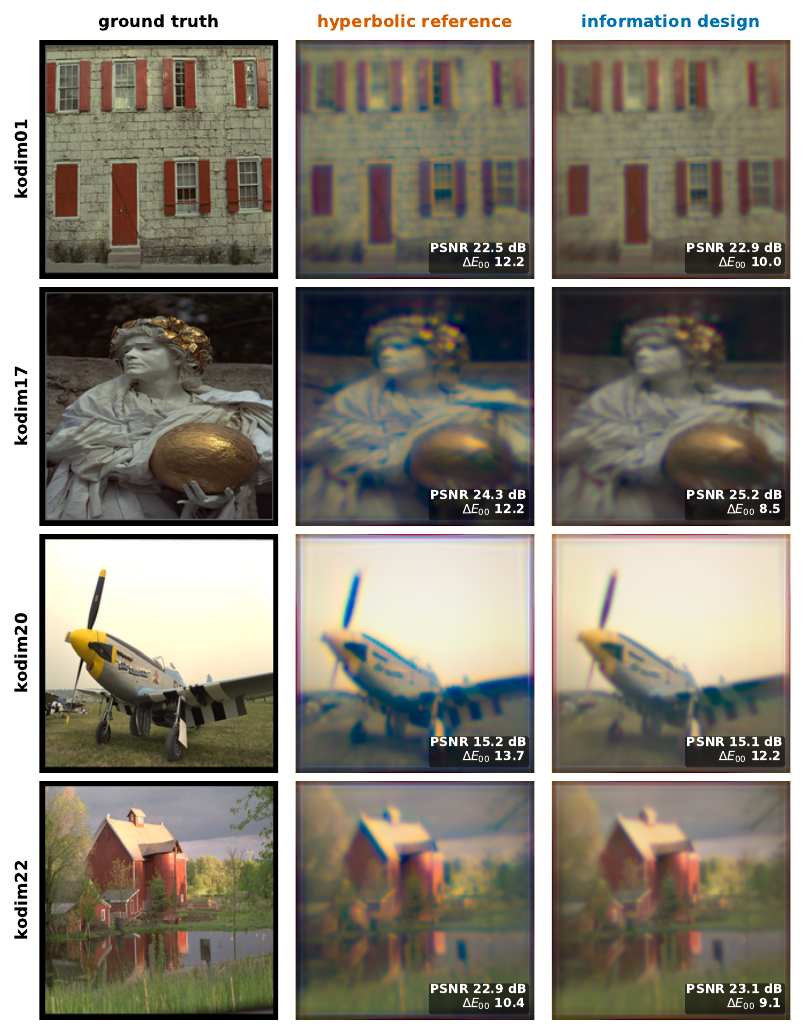}
\caption{\textbf{Reconstructions on four held-out natural scenes (SiN).} Rows,
Kodak test images kodim01, kodim17, kodim20 and kodim22; columns, the ground
truth and the high-SNR reconstructions of the hyperbolic reference and the
information design. Each design is reconstructed under its own independent
analytic color calibration (no per-scene fit), and every image is formed by
direct full-scene propagation with demosaicing and the fixed field-dependent
Wiener step. Each reconstruction carries its PSNR and $\Delta E_{00}$ against the
ground truth (Table~\ref{tab:multiscene}).}
\label{fig:multiscene}
\end{figure}

\begin{figure}[tb]
\centering
\includegraphics[width=0.88\linewidth]{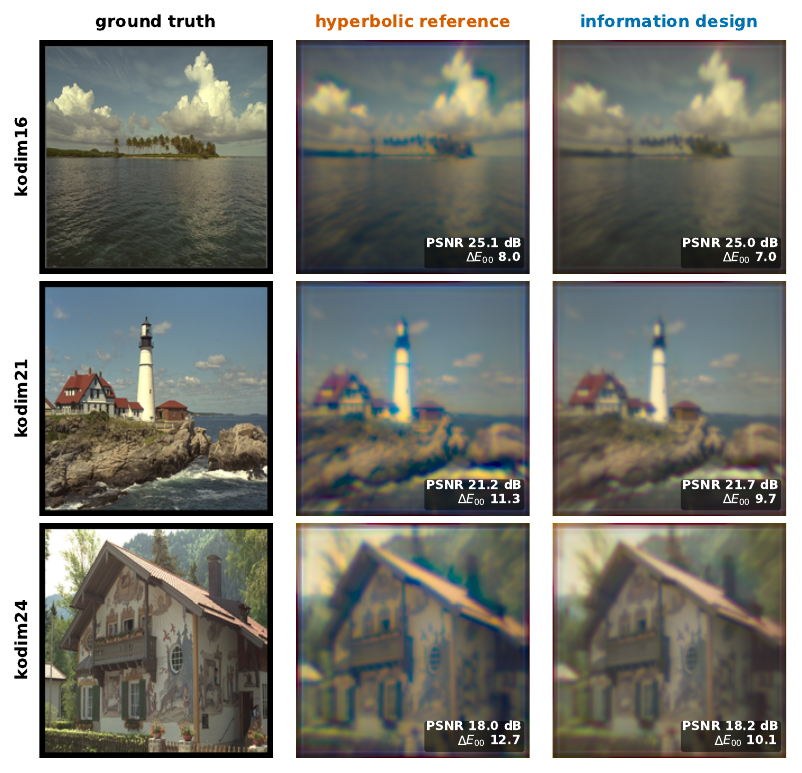}
\caption{\textbf{Reconstructions on three further held-out natural scenes (SiN).}
As Figure~\ref{fig:multiscene}, for the additional Kodak test images kodim16,
kodim21 and kodim24. Each reconstruction carries its PSNR and $\Delta E_{00}$
against the ground truth (Table~\ref{tab:multiscene}).}
\label{fig:multiscene3}
\end{figure}

To check that the imaging advantage is not tied to the single landscape test
image, seven additional held-out natural scenes (Kodak true-color images
kodim01, kodim16, kodim17, kodim20, kodim21, kodim22 and kodim24) were rendered through the identical
pipeline: direct full-scene propagation on the SiN library, demosaicing, the
fixed field-dependent Wiener step, and each design's own independent analytic
color calibration (\S8.3), with no per-scene refitting.
Figures~\ref{fig:multiscene} and~\ref{fig:multiscene3} show the reconstructions and
Table~\ref{tab:multiscene} reports the high-SNR reconstruction metrics. The information design lowers both
color-error measures on all seven scenes and raises PSNR on five of the seven.
The two small PSNR exceptions (kodim16, $-0.05$\,dB and kodim20, $-0.11$\,dB)
still show lower $\Delta E_{00}$ with the information design. Averaged over the
seven scenes the information design lowers $\Delta
E_{00}$ by \num{17.2}\% and S-CIELAB by \num{13.1}\% and raises PSNR by
\SI{0.30}{\decibel}.

\begin{table}[tb]
\centering
\caption{\textbf{Reconstruction metrics on seven held-out natural scenes (SiN).}
High-SNR limit, hyperbolic reference $\rightarrow$ information design, each under
its own independent analytic color calibration (no per-scene fit). PSNR higher
is better. $\Delta E_{00}$ and S-CIELAB lower is better. Bold marks the better
design per scene and metric.}
\label{tab:multiscene}
\begin{tabular}{lccc}
\hline
Scene & PSNR (dB) & $\Delta E_{00}$ & S-CIELAB \\
\hline
kodim01 & 22.48 $\rightarrow$ \textbf{22.89} & 12.19 $\rightarrow$ \textbf{10.01} & 19.88 $\rightarrow$ \textbf{17.26} \\
kodim16 & \textbf{25.06} $\rightarrow$ 25.01 & 7.97 $\rightarrow$ \textbf{7.00}  & 11.13 $\rightarrow$ \textbf{10.22} \\
kodim17 & 24.26 $\rightarrow$ \textbf{25.20} & 12.24 $\rightarrow$ \textbf{8.55}  & 17.45 $\rightarrow$ \textbf{13.08} \\
kodim20 & \textbf{15.21} $\rightarrow$ 15.10 & 13.74 $\rightarrow$ \textbf{12.21} & 25.20 $\rightarrow$ \textbf{22.81} \\
kodim21 & 21.22 $\rightarrow$ \textbf{21.74} & 11.30 $\rightarrow$ \textbf{9.74}  & 18.19 $\rightarrow$ \textbf{16.49} \\
kodim22 & 22.94 $\rightarrow$ \textbf{23.14} & 10.45 $\rightarrow$ \textbf{9.12}  & 16.02 $\rightarrow$ \textbf{14.25} \\
kodim24 & 18.01 $\rightarrow$ \textbf{18.20} & 12.73 $\rightarrow$ \textbf{10.10} & 20.10 $\rightarrow$ \textbf{17.09} \\
\hline
mean    & 21.31 $\rightarrow$ \textbf{21.61} & 11.52 $\rightarrow$ \textbf{9.53}  & 18.28 $\rightarrow$ \textbf{15.89} \\
\hline
\end{tabular}
\end{table}

\subsection*{9.6 Exposure dependence and the concavity prediction}

The objective is concave in modal signal-to-noise, and that concavity is what
makes reallocation profitable (Note 5). The same property predicts where
reallocation stops paying. As the exposure falls, $\log_2(1+\gamma_i)$ approaches
$\gamma_i/\ln 2$, the return on a weak mode matches the return on a strong one,
and the design that collects the most light should recover color best however
that light is distributed. Because the reference collects \num{16}\% more charge
than the information design, the reference is expected to lead in the low-light
limit.

\begin{figure}[tb]
\centering
\includegraphics[width=0.92\linewidth]{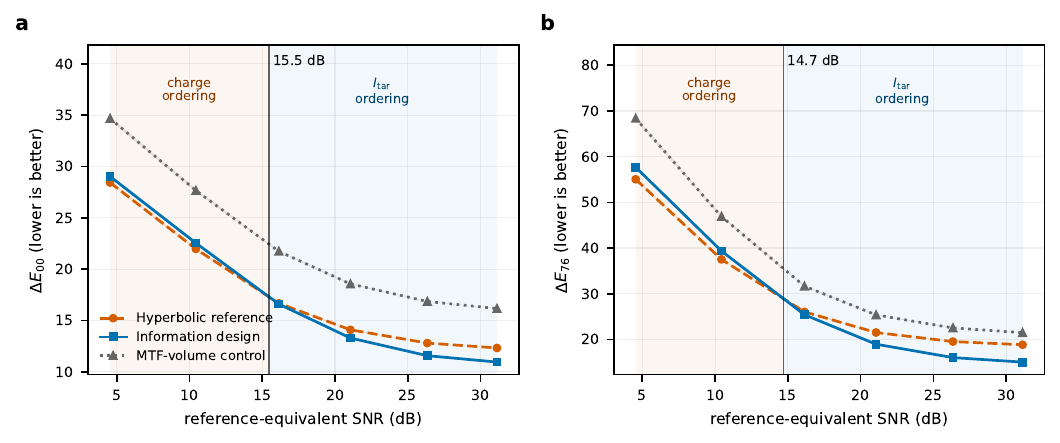}
\caption{\textbf{Exposure dependence of the reconstruction ranking (SiN, single
held-out scene).} \textbf{a}, $\Delta E_{00}$ and \textbf{b}, $\Delta E_{76}$
against reference-equivalent signal-to-noise ratio for the hyperbolic reference,
the information design and the MTF-volume control, each point a mean over 32
paired noise seeds with the $\pm1\sigma$ band. Shading marks the two regimes,
charge-ordered at low exposure and target-information-ordered at high exposure,
separated by the crossing (vertical line). Metrics use each design's own fixed
analytic color calibration. Levels below \SI{0}{\decibel}, under one electron
per pixel, are omitted from the plot.}
\label{fig:noise}
\end{figure}

The sweep follows that prediction. The held-out landscape was reconstructed at
eight common-exposure levels spanning reference-equivalent signal-to-noise ratios
from \SI{-11.6}{\decibel} to \SI{31.1}{\decibel}, with 32 paired noise seeds at
each level and the color calibration and reconstruction of \S8 held fixed
throughout (Fig.~\ref{fig:noise}). At the four dimmest levels both color measures
rank the three designs by collected charge, placing the hyperbolic reference
first at \num{724.0} electrons per raw pixel, the information design second at
\num{624.4}, and the MTF-volume control last at \num{274.8}. At the brightest
levels both measures rank them instead by delivered target information, and the
information design leads. The crossing falls at \SI{15.5}{\decibel} for
$\Delta E_{00}$ and \SI{14.7}{\decibel} for $\Delta E_{76}$, where the mean signal
is about \num{40} electrons per pixel. Each noise realization is shared across the
three designs, so the ranking is resolved seed by seed everywhere except within
about \SI{1}{\decibel} of the crossing. Above \SI{21}{\decibel} the information
design leads in $\Delta E_{00}$ by more than fifteen times the paired seed spread.
PSNR does not resolve the transition and is omitted.

The crossing sits in the lower-exposure part of the swept range. The bright end
of the sweep is bounded by the modeled full well of
\num{12000} model-electrons, where saturation begins near
\num{3900} model-electrons per pixel and caps the attainable ratio at about
\SI{41}{\decibel}. These are calibrated model-electron levels, not an absolute
radiometric mapping to scene illuminance.
Reallocating transfer beats collecting more of it once the
measurement is strong enough for the concavity to matter, and the sweep locates
that threshold for this platform and scene. These are descriptive numbers for a
single scene.

% =====================================================================
\snote{10: Stated limitations}
\label{sn:limits}

\medskip\noindent\textbf{This is a computational study.} No device was
fabricated and no optical measurement was made. Every quantity is from a
simulated forward model. Fabrication and paired tolerance analysis, including
width bias, height error, sidewall angle and spatially correlated process
variation, are the immediate next steps.

\medskip\noindent\textbf{The forward model is local-periodic.} Each post is
assigned the response of an infinite array of identical posts, tabulated at
Fourier order $(7,7)$ on a $256^2$ raster with C4 completion. We tested this
approximation directly on the full target-information width map.

\medskip\noindent\textbf{Exact-map Tidy3D FDTD diagnostic.} The complete
$D=208$\,\um{} map contains 404{,}048 silicon-nitride square posts on the
290\,nm lattice, with 1.0\,\um{} height and the optimized widths left unchanged.
All posts were represented explicitly in a Tidy3D finite-difference time-domain
calculation. A substrate-side, normally incident, $x$-polarized broadband pulse
covered the nine design wavelengths
\SIlist{420;450;470;510;540;570;600;635;670}{nm}. The $x$ and $y$ grid spacing
was 20\,nm, the $z$ grid used at least 14 steps per wavelength, and the two exact
Cartesian mirror symmetries of the stored map reduced the simulated volume. The
complex $E_x$, $E_y$ and $E_z$ fields were recorded at $z=1.35$\,\um{} and
propagated through air to the same 140 longitudinal planes used by the cached
local-periodic simulator. The same centered $D=208$\,\um{} circular aperture was
applied to the complex Tidy3D monitor fields before propagation, matching the
simulator pupil.

For each wavelength, both $x$--$z$ intensity maps were interpolated to the same
720-point transverse grid and normalized to unit integrated intensity. This
removes an arbitrary wavelength-dependent gain and tests the spatial field
shape. Across all nine wavelengths the mean total-variation distance is
\num{0.0892}, with range \num{0.0743} to \num{0.1217}, and the mean cosine
similarity is \num{0.9863}, with range \num{0.9674} to \num{0.9931}. Seven focal
positions agree on the common grid. The other two differ by one longitudinal
sample, \SI{2.73}{\um}, giving an rms focal-position difference of
\SI{1.29}{\um} and mean signed difference of \SI{0.61}{\um}.

We also compared the transverse point-spread functions on a common plane, using
the simulator peak position $z^{*}(\lambda)$ for both models at each wavelength. The
PSFs were interpolated to the same grid, restricted to
$|x|,|y|\leq\SI{15}{\um}$ and normalized to unit integrated intensity. Across
the nine wavelengths the mean PSF total-variation distance is \num{0.0644},
with range \num{0.0486} to \num{0.0900}, and the mean PSF cosine similarity is
\num{0.9990}, with range \num{0.9984} to \num{0.9993}.

The comparison is on axis and uses intensity fields from a finite 0.50\,ps run,
which ended with normalized field decay $4.78\times10^{-5}$. It establishes
agreement of the longitudinal and transverse intensity, not of the full complex
vector field or the off-axis response.

\begin{figure}[htbp]
\centering
\includegraphics[width=\linewidth]{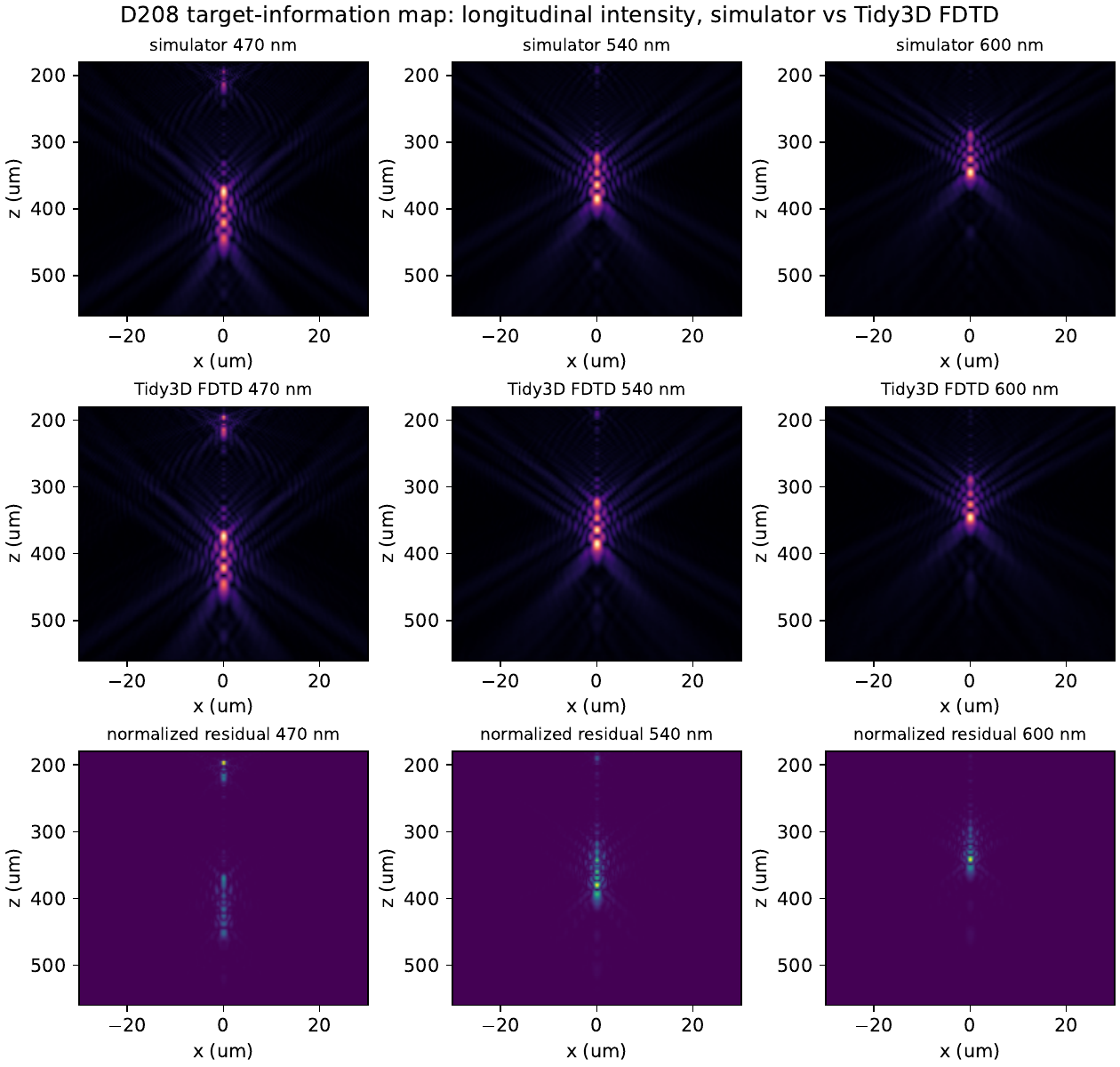}
\caption{\textbf{Exact-map Tidy3D FDTD $x$--$z$ comparison.} The cached
local-periodic simulator (top) and Tidy3D FDTD calculation (middle) for the
exact $D=208$\,\um{} target-information map at
\SIlist{470;540;600}{nm}. Each panel is normalized by its own peak for display.
The same centered $D=208$\,\um{} circular pupil is used for both models.
The bottom row shows the absolute residual after unit-mass normalization. The
comparison covers the complete longitudinal intensity map rather than only the
focal position. Table~\ref{tab:fdtd} reports the all-wavelength statistics.}
\label{sfig:fdtd}
\end{figure}

\begin{figure}[htbp]
\centering
\includegraphics[width=\linewidth]{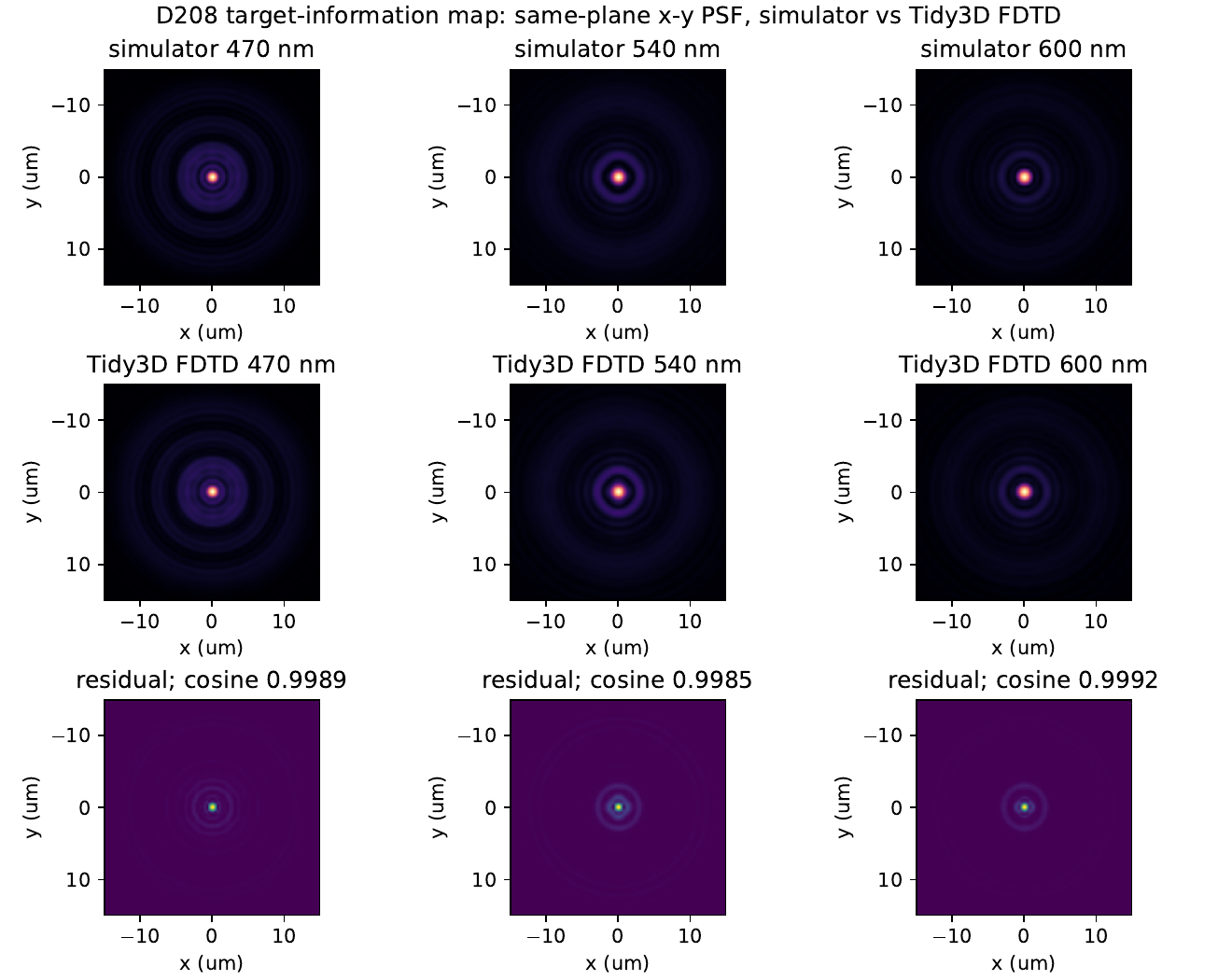}
\caption{\textbf{Exact-map Tidy3D FDTD $x$--$y$ PSF comparison.} The cached
local-periodic simulator (top) and Tidy3D FDTD calculation (middle) for the
exact $D=208$\,\um{} target-information map at
\SIlist{470;540;600}{nm}. Both models are evaluated at the same
simulator-predicted $z^{*}(\lambda)$ defined in the text, on a common
$\SI{30}{\um}\times\SI{30}{\um}$ transverse window and use the same centered
$D=208$\,\um{} circular pupil. Each PSF is normalized by
its own peak for display. The bottom row shows the absolute residual after
unit-mass normalization and states the corresponding cosine similarity.}
\label{sfig:fdtd-psf}
\end{figure}

\begin{table}[htbp]
\centering
\caption{\textbf{Exact-map Tidy3D FDTD diagnostic.} Scale-invariant spatial
agreement between the cached local-periodic simulator and Tidy3D FDTD for the exact
$D=208$\,\um{} target-information map. The reported range is over all nine
design wavelengths. The $x$--$y$ PSFs use the same simulator-predicted
$z^{*}(\lambda)$ for both models.}
\label{tab:fdtd}
\begin{tabular}{lll}
\toprule
Observable & Aggregate value & Wavelength range\\
\midrule
$x$--$z$ total-variation distance & \num{0.0892} mean & \num{0.0743}--\num{0.1217}\\
$x$--$z$ cosine similarity & \num{0.9863} mean & \num{0.9674}--\num{0.9931}\\
$x$--$y$ PSF total-variation distance & \num{0.0644} mean & \num{0.0486}--\num{0.0900}\\
$x$--$y$ PSF cosine similarity & \num{0.9990} mean & \num{0.9984}--\num{0.9993}\\
Focal-position difference & \SI{0.61}{\um} mean, \SI{1.29}{\um} rms & \SIrange{0}{2.73}{\um}\\
\bottomrule
\end{tabular}
\end{table}

\medskip\noindent\textbf{The objective is exact only within the stated
model.} $\mathcal I_{\mathrm{tar}}$ is a channel--target quantity whose
determinant form uses conditional-mean sufficiency under the linear--Gaussian
model with one empirical scene prior and one fixed exposure. A learned
nonlinear decoder under non-Gaussian statistics could favour a different
allocation.

% =====================================================================
\bibliography{combined_refs}

\end{document}